\documentclass[useAMS,usegraphicx,usenatbib]{mn2e}

\usepackage{graphicx}
\usepackage{times}
\usepackage{amssymb}
\usepackage{amsmath}
\usepackage{euscript}
\usepackage{longtable,lscape}

\def\210keV{{\rm\thinspace 2--10 keV}}
\usepackage{wasysym}

\title[Bolometric Corrections for AGN]{Piecing Together the X-ray Background: Bolometric Corrections for Active Galactic Nuclei}
\author[R.V. Vasudevan \& A.C. Fabian]{R.V. Vasudevan$^1$\thanks{e-mail:ranjan@ast.cam.ac.uk} and A.C. Fabian$^1$\\\footnotesize$^1$ Institute of Astronomy, Madingley Road, Cambridge CB3 0HA}

\begin{document}

\maketitle

\begin{abstract}
The X-ray background can be used to constrain the accretion history of Supermassive Black Holes (SMBHs) in Active Galactic Nuclei (AGN), with the SMBH mass density related to the energy density due to accretion. A knowledge of the hard X-ray bolometric correction, $\kappa_{\mathrm{2-10keV}}$ is a vital input into these studies, as it allows us to constrain the parameters of the accretion responsible for SMBH growth.  Earlier studies assumed a constant bolometric correction for all AGN, and more recent work has suggested accounting for a dependence on AGN luminosity. Until recently, the variations in the disk emission in the UV have not been taken into account in this calculation; we show that such variations are important by construction of optical -- to -- X-ray SEDs for 54 AGN. In particular, we use FUSE UV and X-ray data from the literature to constrain the disk emission as well as possible.  We find evidence for very significant spread in the bolometric corrections, with no simple dependence on luminosity being evident.  Populations of AGN such as Narrow-Line Seyfert 1 nuclei (NLS1s), Radio Loud and X-ray Weak AGN may have bolometric corrections which differ systematically from the rest of the AGN population.  We identify other sources of uncertainty including intrinsic extinction in the optical--UV, X-ray and UV variability and uncertainties in SMBH mass estimates.  Our results suggest a more well-defined relationship between bolometric correction and Eddington ratio in AGN, with a transitional region at an Eddington ratio of $\sim0.1$, below which the bolometric correction is typically $15 - 25$, and above which it is typically $40 - 70$.  We consider the potential implied parallels with the low/hard and high/soft states in Galactic Black Hole (GBH) accretion, and present bolometric corrections for the GBH binary GX 339-4 for comparison. Our findings reinforce previous studies proposing a multi-state description of AGN accretion analogous to that for GBH binaries.  Future calculations of the SMBH mass density may need to take into account the possible dependence of $\kappa_{\mathrm{2-10keV}}$ on Eddington ratio.

\end{abstract}


\section{Introduction}\label{Intro}

Numerous authors have investigated the relationship between past Active Galactic Nuclei (AGN) activity and the local supermassive black hole (SMBH) population.  A detailed study of the relationship between the two can yield important insights into the nature of AGN activity, providing us with parameters such as the typical accretion efficiency and Eddington ratio for AGN.

Early work by  \cite{1982MNRAS.200..115S} demonstrated how the total mass density of SMBHs in the local universe could be inferred from counts of quasars/AGN (hereafter referred to as the `Soltan argument'). \cite{2004MNRAS.354.1020S} and \cite{2004MNRAS.351..169M} demonstrate how the mass function of local SMBHs can be matched with that expected from accreting AGN, for appropriately chosen accretion efficiencies and Eddington ratios.

Various observational studies have convincingly established that the X-ray Background (XRB) can be resolved as the sum of the spectra from AGN (e.g. as summarised in \citealp{2004NuPhS.132...86H}).  It is then possible to use the X-ray background light as a `signature' for accretion causing SMBH growth, and so the energy density in the XRB can be converted into a mass density of SMBHs expected in the present epoch, taking into account the efficiency of SMBH accretion in the process.  The basic method for this is summarised by \cite{Fabian:2003cz}:  if the observed energy density due to accretion is given by $U_{T}^*$ and the mass--to--radiation conversion efficiency is $\eta$, then the local SMBH mass density $\rho_{\mathrm{BH}}$ is:

\begin{equation}
\label{eq1}
\eta\rho_{\mathrm{BH}}c^2 = (1+{\langle}z{\rangle}) U_{\mathrm{T}}^*\\
\end{equation}
where ${\langle}z{\rangle}$ is the mean redshift of accreting AGN.

However, we require a bolometric correction $\kappa_{\mathrm{2-10keV}}$ to scale up the 2-10 keV XRB spectral intensity to obtain the total accretion energy density:
\begin{equation}
\label{eq2}
 U_{\mathrm{T}}^* = \frac{4\pi}{c} \kappa_{\mathrm{2-10keV}} I_{\mathrm{XRB}}.
\end{equation}
where $I_{\mathrm{XRB}}$ is the \210keV XRB spectral intensity.  The quantity $U_{T}^*$ can then be used in equation (\ref{eq1}) to determine the SMBH mass density.  Despite previous detailed studies into AGN and quasar spectral energy distributions (SEDs) such as the seminal work by \cite{1994ApJS...95....1E}, there remain significant issues regarding the determination of this bolometric correction.  This can have wide-ranging effects, since the choice of bolometric correction will directly constrain the allowed efficiencies for which the Soltan argument can be satisfied.  The efficiency of accretion is directly related to black hole spin, and so the quantity $\kappa_{\mathrm{2-10keV}}$ needs to be well understood.

Previously, bolometric corrections have been determined from a Mean Energy Distribution (MED) for AGN calculated from 47 quasars \citep{1994ApJS...95....1E}.  It is important to note that the infrared (IR) emission is known to be re-processed from the ultraviolet (UV), as discussed by \cite{1993ARA&A..31..473A}.  The bolometric corrections of \cite{1994ApJS...95....1E}, which include the IR emission in the bolometric luminosity, effectively count part of the emission twice and the resulting bolometric corrections would be too high \citep{2004MNRAS.351..169M}, and would overestimate the accretion luminosity.  Another key issue is the configuration of the absorbing material responsible for producing the IR emission: while the UV and X-ray emission from the disk and corona are thought to be emitted more isotropically, the IR emitting material is unlikely to cover the source completely.  Therefore in the process of converting the IR flux into a luminosity, the implicit assumption of isotropic emission would yield an artificially high IR contribution.  To recover the \emph{intrinsic} accretion luminosity in AGN (which emerges in the optical, UV and X-ray bands), in general the IR emission should be excluded when calculating bolometric corrections.  There could be an important synchrotron contribution to the IR in Radio Loud AGN as discussed in \cite{2004MNRAS.351..169M}, but we do not explore this further here (only around 10 per cent of our objects are radio loud).  Additionally, \cite{1994ApJS...95....1E} themselves caution against using the MED to determine global properties of quasars and AGN, since the individual SEDs used to determine the MED themselves show significant variation over many wavelength bands.

More recent studies have accounted for variations in bolometric correction using a re-scalable template SED. \cite{2004MNRAS.351..169M} account for variations in AGN SEDs by using the well known anti-correlation between the optical--to--X-ray spectral index ($\mathrm{\alpha_{OX}}$) and $\mathrm{2500\AA}$ luminosity \citep{2003AJ....125..433V} to construct such a template, where $\mathrm{\alpha_{OX}}$ is defined as:

\begin{equation}
\label{eq3}
\mathrm{\alpha_{OX}} = -\frac{\mathrm{log[L_{\nu}(2500\AA)/L_{\nu}(2keV)]}}{\mathrm{log[\nu(2500\AA)/\nu(2keV)]}}. 
\end{equation}

\begin{figure}
    \includegraphics[width=7cm]{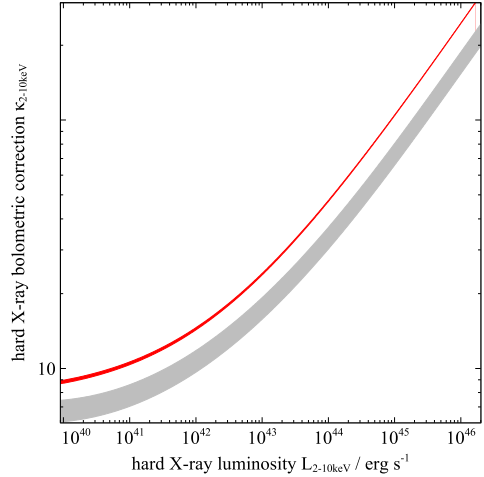}
    \caption{Bolometric correction as a function of luminosity, with  spread calculated from Monte-Carlo realisations of a template AGN SED, from \protect\cite{2004MNRAS.351..169M} (lower grey filled area) and \protect\cite{2006astro.ph..5678H} (higher red filled area).}
\label{fig:marconibolcor}
\end{figure}

\cite{2004MNRAS.351..169M} renormalise their template SED to a particular value of $\alpha_{OX}$ to obtain the variation of bolometric correction with luminosity.  A similar approach is followed by \cite{2006astro.ph..5678H}, but employs a template SED generated from the averages of real SEDs in different wavebands. The variations of bolometric correction with luminosity calculated by both of these studies are reproduced in Fig. \ref{fig:marconibolcor}. The latter paper does advise that intrinsic spread in AGN SEDs could give rise to variation of a factor of $\sim2$ in the bolometric correction, even when luminosity dependence is taken into account.

We have constructed basic SEDs for 54 AGN and calculated their bolometric corrections, for comparison with the trends suggested by these authors.  Historically it has been established that the optical to X-ray SED of AGN can be relatively well described by a black body disk spectrum peaking in the UV and a power-law model extending to hard X-rays; but most emission from lower luminosity objects is unobserved by virtue of it peaking in the absorbed far ultra-violet (FUV).  We therefore used UV observations as the foundation for our study, to constrain the disk emission component.  We employed data from a range of sources for optical and X-ray SED points and fitted all the available multi-wavelength data to a basic disk and power-law model.

The comparison between accretion on large and small scales is a field of much interest, and recent work has sought to describe accretion across all mass scales in terms of similar physical processes and accretion states (e.g. \citealt{2006Natur.444..730M}).  The accretion states in Galactic Black Hole binaries (GBHs) are seen to be separated by different Eddington ratios ($\mathrm{L_{bol}/L_{Edd}}$), and to facilitate comparison with AGN we present the bolometric corrections as a function of Eddington ratio.  We compare our findings qualitatively to the current understanding of state transitions in GBH X-ray binaries.

\section{Data sources}
\label{sec:datasources}

The starting point of our study is the sample of 85 low redshift ($z < 0.7$) AGN surveyed by the \emph{Far Ultraviolet Spectroscopic Explorer} (\emph{FUSE}) as presented by \cite{2004ApJ...615..135S}; this provides a measure of the ``big blue bump'' (BBB) emission from the thermal accretion disk in the UV.  Observations from \emph{HST-STIS} have been used where \emph{FUSE} data were not available.  Optical and X-ray data have been gathered from publications in the literature employing a variety of instruments, including the Hubble Space Telescope (\emph{HST}), Kitt Peak National Observatory (\emph{KPNO}), the Advanced Satellite for Cosmology and Astrophysics (\emph{ASCA}, as presented in the \textsc{Tartarus} Database\footnote{http://astro.ic.ac.uk/Research/Tartarus/}) and \emph{XMM-Newton}. Complete datasets (UV, X-ray and optical if available) were obtained for 54 AGN of which 48 were drawn from the \emph{FUSE} sample of Scott et al.  The data sources used are listed in Table \ref{table:datasources}1 in Appendix \ref{appendix1}.

\subsection{Sample Selection}
\label{sampleselection}

The study of \cite{1994ApJS...95....1E} is consists primarily of X-ray bright AGN drawn from an X-ray selected sample.Therefore an X-ray bolometric correction calculated from their mean SED would be an underestimate for the AGN with more canonical X-ray luminositites. Since our sample draws largely from the AGN sample presented in \cite{2004ApJ...615..135S}, any selection biases present in their sample will necessarily transfer directly to our study.  By nature, the \emph{FUSE} study will be UV bright and indeed contains few very low luminosity ($\mathrm{L_{X}} < 10^{43} \mathrm{erg s^{-1}}$) Seyferts, and so for the lowest luminosity objects in table \ref{table:datasources}1 we obtain UV data from other sources. The preponderance for brighter Seyferts and quasars should be noted at the outset.

Prior to the work of \cite{2004ApJ...615..135S}, \cite{2002ApJ...565..773T} provided an analysis of the UV spectral properties of 184 QSOs with UV observations from \emph{HST}.  They found a distinctly softer average Extreme-UV (EUV) spectral slope ($\mathrm{\alpha_{EUV}} = -1.76 \pm 0.12$ compared to Scott et al.'s $\mathrm{\alpha_{EUV}} = -0.56^{+0.38}_{-0.28}$ for $f_{\nu} \propto {\nu}^{\alpha_{EUV}}$).  Their sample is brighter in the UV than the \cite{2004ApJ...615..135S} sample, with an average value of $\mathrm{log(\lambda L_{1100\AA})}\sim 45.9$ compared to $\sim 45.0$ for the \cite{2004ApJ...615..135S} sample.  Additionally, the median redshift of the \cite{2002ApJ...565..773T} sample is $0.96$ compared to $0.1$ for the \cite{2004ApJ...615..135S} sample.  By cosmic downsizing, it is reasonable to expect significantly higher SMBH masses for the higher redshift Telfer et al. sample.  This systematic difference in the average mass could account for the difference in spectral shape, possibly reducing the fraction of luminosity from the BBB.  If the SEDs from the Telfer et al. sample were assumed to lie on the $\mathrm{\alpha_{OX}}$ -- $\mathrm{2500\AA}$ luminosity relation, we would expect similar bolometric corrections, with the possibility of some reduction due to the smaller disk fraction extrapolated from the softer EUV slope.  

Without a record of the objects in the \cite{2002ApJ...565..773T} sample along with mass estimates (to constrain the shape of the disk model in the UV), it is difficult to make any predictions as to how the bolometric corrections might vary between these two samples. We concentrate on the lower redshift objects of interest here, as presented in the \emph{FUSE} sample.


When attempting to put together this sample, the prime concern was to assemble a catalogue of objects for which at least UV, X-ray and mass data were available and so the sample selection has consequently been constrained by the availability of published data on the sources of interest.  However the subsequent analysis reveals that despite the biases of the parent samples, there are a range of object classes represented, notably both radio loud and radio quiet AGN as well as a substantial number of Narrow Line Seyfert 1 nuclei (NLS1s).

\section{Construction of SEDs}

The multi-wavelength data points were converted into \textsc{XSPEC} spectra files using the \textsc{flx2xsp} utility, part of the \textsc{ftools} package.  The data points (corrected for redshift) were fit to the \textsc{diskpn} model in the \textsc{xpsec} package, with a simple power law at X-ray energies up to $\mathrm{250keV}$ (\textsc{powerlaw} model) or, if the X-ray photon index was steeply decreasing, a broken power law increasing steeply within the UV disk region (\textsc{bknpower} model) in order to prevent the extrapolation of the powerlaw from producing artificially high fluxes at low energies.  The normalisation of the disk component was constrained using SMBH mass estimates available in the literature and was kept frozen during fitting.  The disk temperature was seeded with a typical value of $\mathrm{0.01keV}$ and subsequently allowed to vary, and the inner radius of the accretion disk was frozen at $\mathrm{6 GM/c^2}$; appropriate for the high accretion efficiencies found from previous applications of the Soltan argument.

If multiple epoch X-ray observations were available, the mean $2-10\mathrm{keV}$ flux and photon index from the available observations were used.  The \textsc{Tartarus} database provides both an `absorbed' and an `unabsorbed' model fit to the ASCA X-ray data, and if the fit parameters ($F_{2-10\mathrm{keV}}$, \210keV flux and photon index $\Gamma$) were significantly different between the two fits, the fit parameters with lower reduced chi-squared were used.  If both had similar reduced chi-squared values but different model fit parameters, the physically simpler `unabsorbed' parameters were used.  Where errors were not provided (e.g. on individual fluxes and photon indices in the \textsc{Tartarus} database), errors of $\pm 0.1$ on spectral indices and $\pm 10$ per cent on fluxes were assumed.  Optical points were taken primarily from the study of \cite{2005MNRAS.356.1029B} and NASA/IPAC Extragalactic Database (NED) photometry. The optical points did not always follow the shape implied by a BBB expected in the optical--UV.  When the optical fluxes lay significantly above the UV points, deviating markedly from the best disk model fit, they were not included in a revised SED fit.  This optical excess may be due to host galaxy starlight contamination, or significant reddening in the UV causing a real decrease in the UV fluxes with respect to the optical fluxes.  In this study we make the assumption that the AGN have at least a basic BBB shape, in order to fit a disk model to the UV-optical region.

\begin{figure}
   \includegraphics[width=4.3cm]{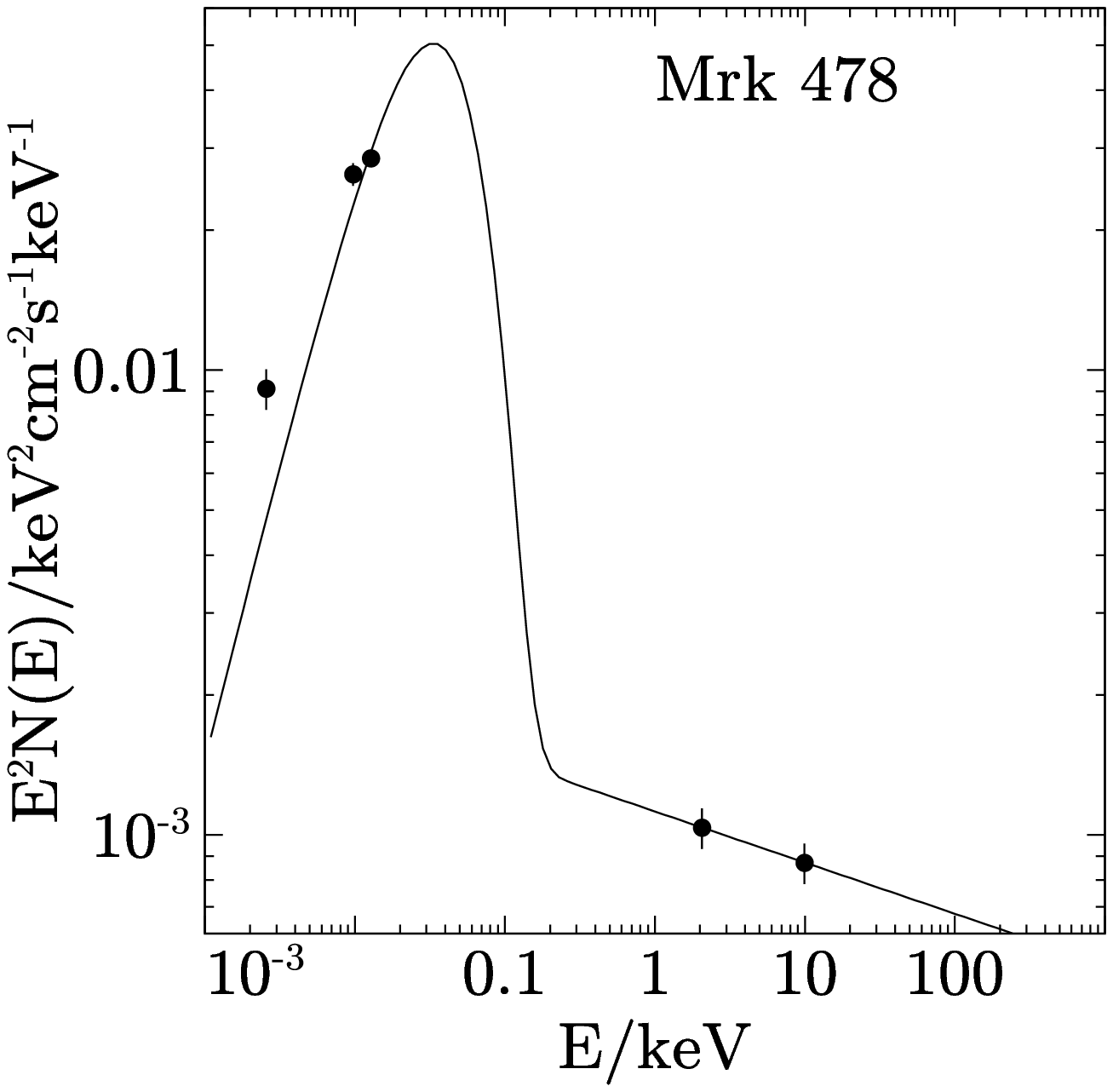}
   \includegraphics[width=4.3cm]{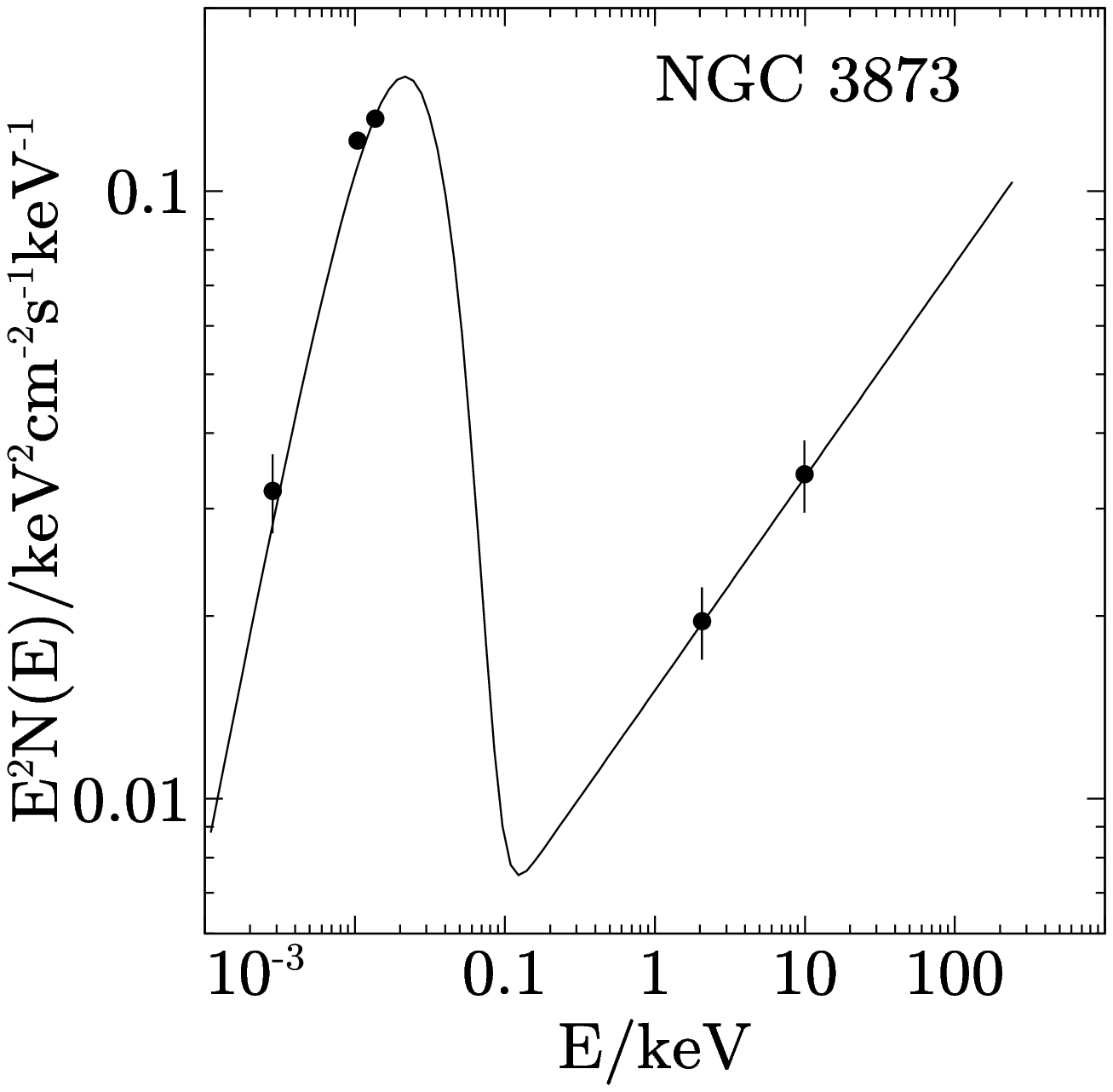}
   \includegraphics[width=4.3cm]{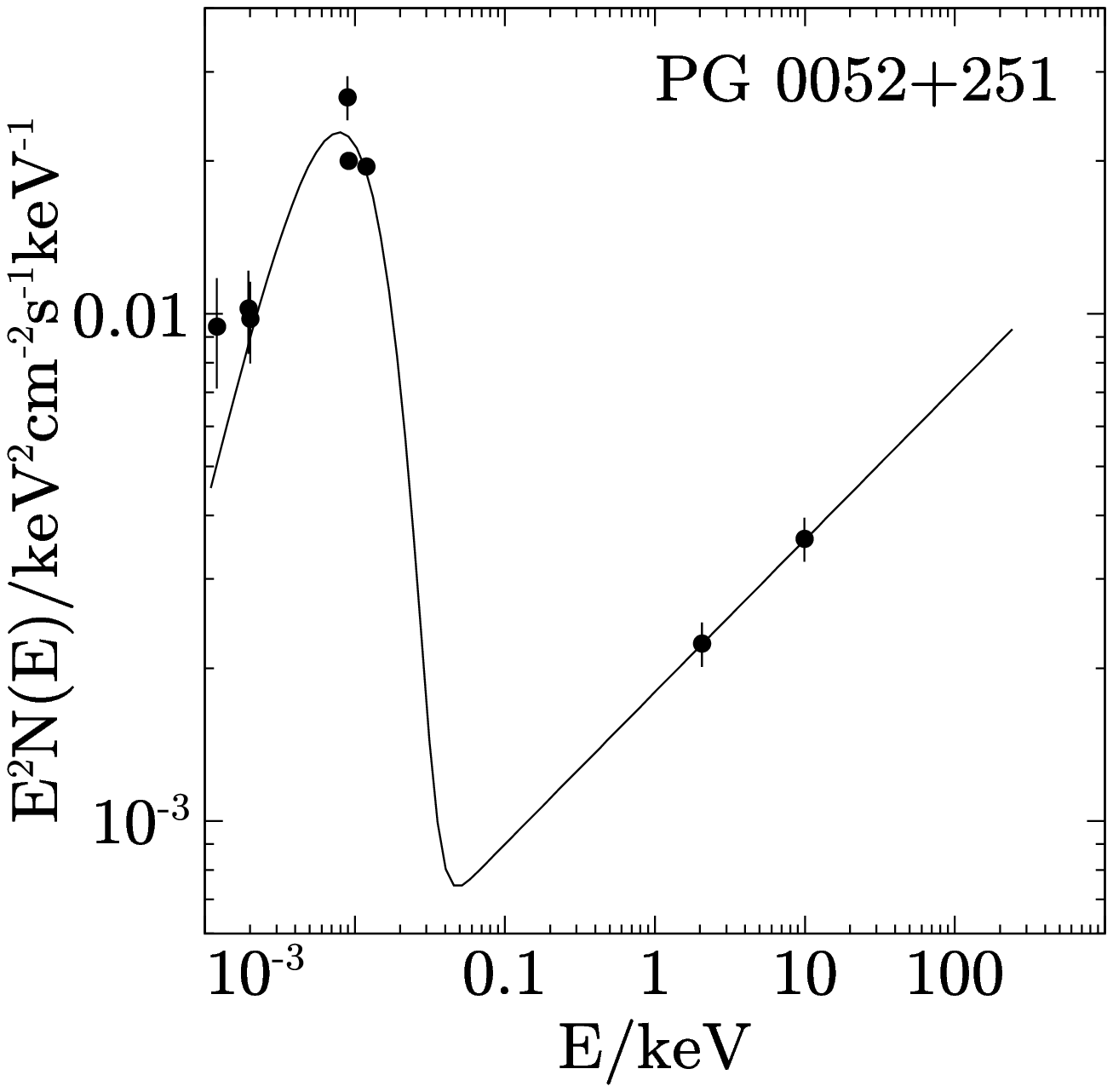}
   \includegraphics[width=4.3cm]{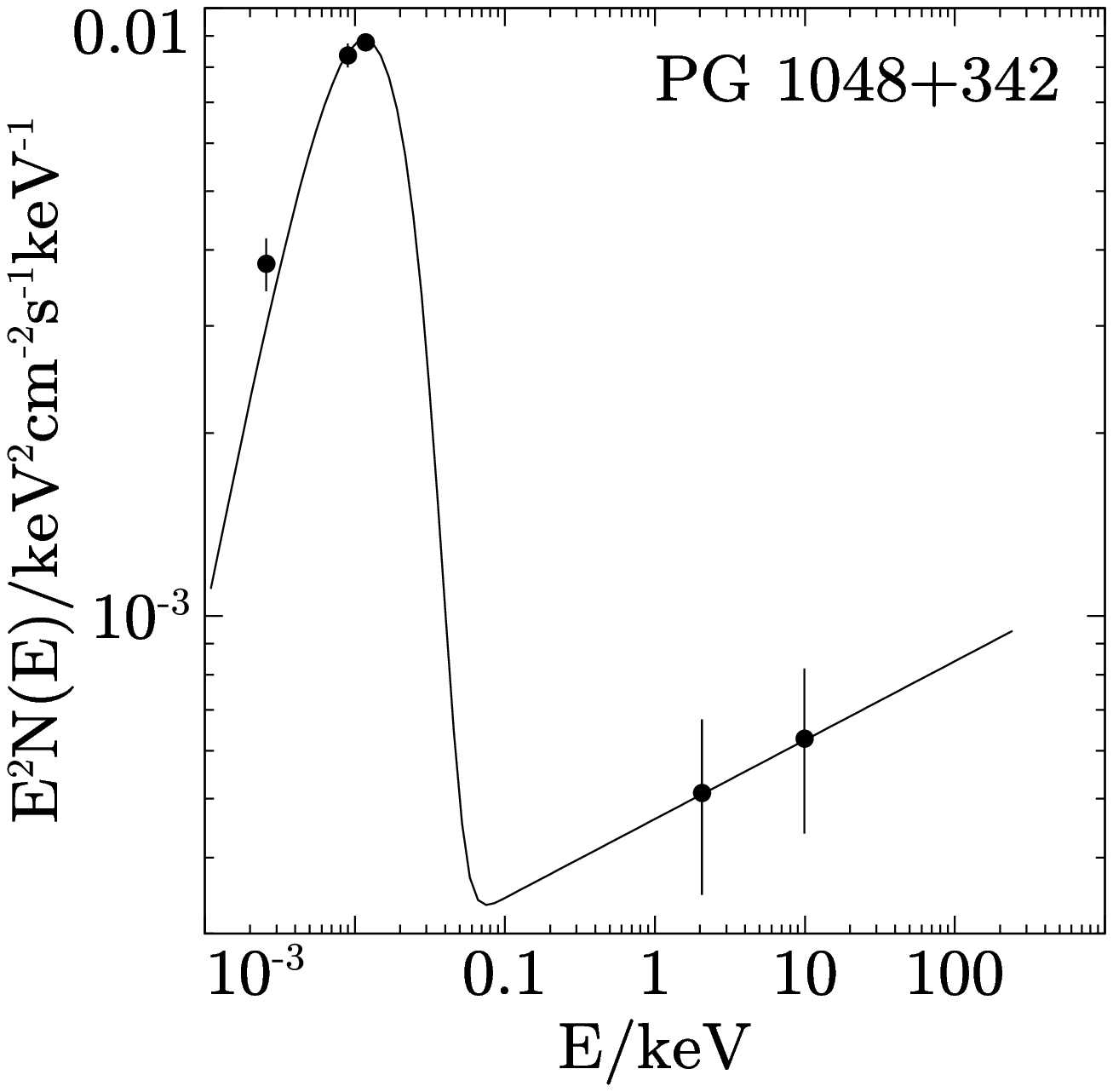}
   \caption{Some sample SEDs.  Filled circles are the data points obtained from the sources detailed in section \ref{sec:datasources} and solid lines show the model fits.}
\label{sampleSEDs}
\end{figure}

\begin{figure*}
   \includegraphics[width=14cm]{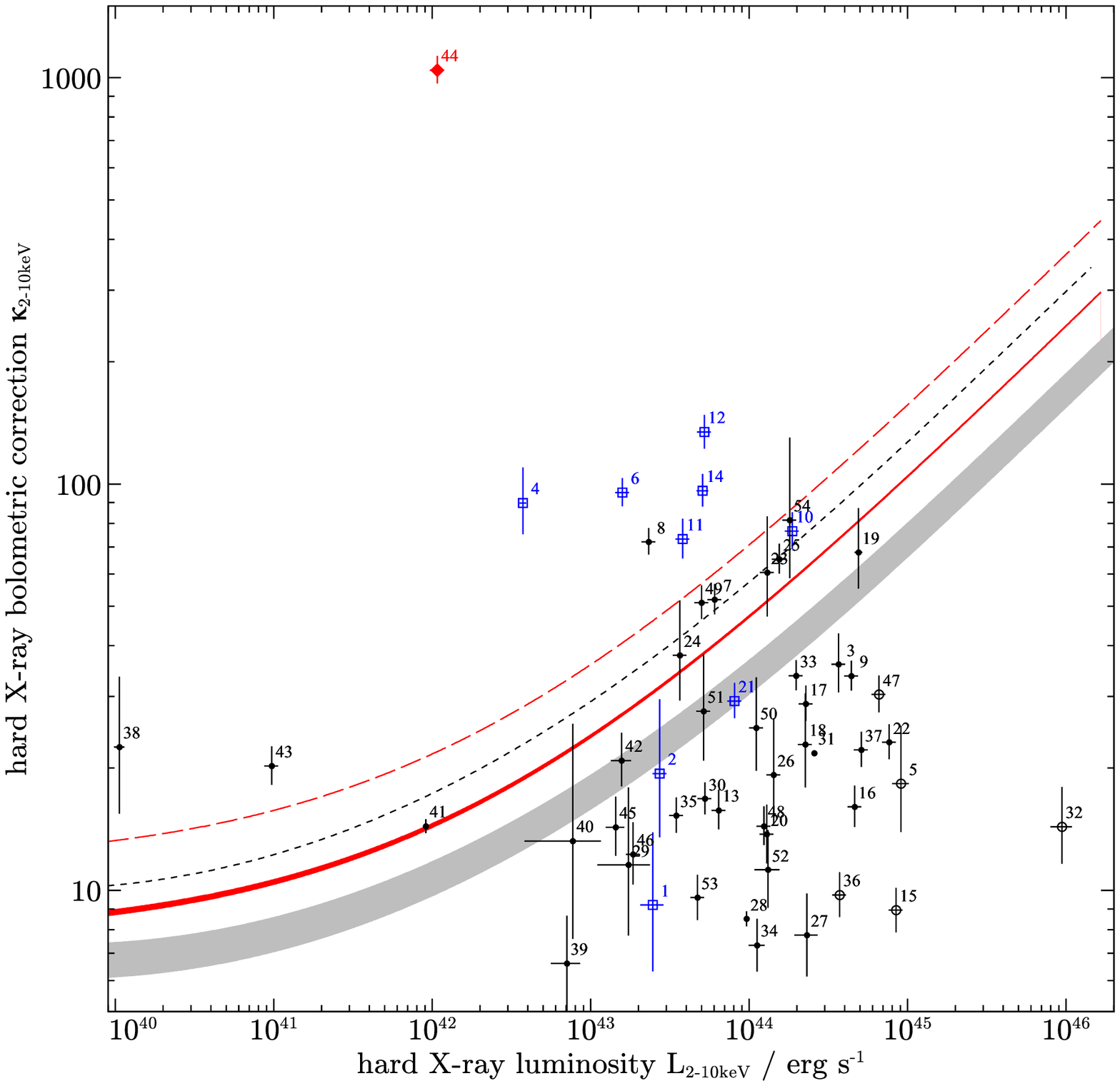}
   \caption{Hard X-ray Bolometric Correction against \210keV luminosity for all AGN in sample. The grey (lower) and red (higher) shaded areas represent the bolometric corrections (with spread) of \protect\cite{2004MNRAS.351..169M} and \protect\cite{2006astro.ph..5678H} respectively.  The dashed black and red lines represent the bolometric corrections that would have been obtained by \protect\cite{2004MNRAS.351..169M} and \protect\cite{2006astro.ph..5678H} if the IR had been included at a level of $\sim 1/3$ of the total luminosity.  Blue empty squares represent Narrow Line Seyfert 1 nuclei (NLS1s), black empty circles represent Radio Loud AGN (RL) and red diamonds represent X-ray weak.  The AGN in the sample: 1 - Ark 564, 2 - I Zw 1, 3 - PG 1116+215, 4 - KUG 1031+398, 5 - PG 1512+370, 6 - Mrk 335, 7 - PG 1322+659, 8 - TON 951, 9 - PG 0026+129, 10 - PG 1402+261, 11 - Mrk 478, 12 - PG 1211+143, 13 - Ark 120, 14 - TONS 180, 15 - 4C +34.47, 16 - PG 0052+251, 17 - PG 0804+761, 18 - PG0947+396, 19 - PG 0953+414, 20 - PG 1114+445, 21 - PG 1115+407, 22 - PG 1216+069, 23 - PG 1307+085, 24 - PG 1415+451, 25 - PG 1444+407, 26 - PG 1626+554, 27 - PG 1229+204, 28 - Mrk 205, 29 - Mrk 79, 30 - Mrk 279, 31 - MR 2251-178, 32 - 3C 273, 33 - Mrk 1383, 34 - Fairall 9, 35 - NGC 985, 36 - 3C 382, 37 - PG 2349-014, 38 - NGC 4395, 39 - NGC 4593, 40 - NGC 3516, 41 - NGC 3227*, 42 - NGC 7469, 43 - NGC 6814, 44 - PG 1011-040, 45 - NGC 3873, 46 - Mrk 290, 47 - PG 1100+772, 48 - PG 1352+183, 49 - ESO 141-G55, 50 - PG 1048+342, 51 - II Zw 136, 52 - Mrk 509, 53 - Mrk 506, 54 - Mrk 876.  * De-reddened UV fluxes assumed for NGC 3227 to allow a fit to the disk model - see \protect\cite{2001ApJ...555..633C} for discussion.}
\label{bcvslumallagn}
\end{figure*}

The hard X-ray bolometric correction for each AGN was determined by dividing the luminosity in the energy range $\mathrm{0.001-250keV}$ (i.e. from optical wavelengths to hard X-rays) by the \210keV luminosity.  Some sample SEDs generated via this process are shown in Fig. \ref{sampleSEDs}, with bolometric correction plotted against luminosity for the 54 AGN in Fig. \ref{bcvslumallagn}. The bolometric corrections for the AGN in the sample are reproduced in Table \ref{table:bolcorstable}.

\begin{table*}
\begin{tabular}{p{2cm}p{1.3cm}p{2.8cm}p{2.7cm}p{2.8cm}p{2.0cm}p{1.5cm}}
\hline
AGN&Redshift&$\mathrm{L_{2-10keV}}$ ($\mathrm{10^{44}erg s^{-1}}$) [1]&\begin{sloppypar}$\mathrm{L_{bol}}$ ($\mathrm{10^{44}erg s^{-1}}$)\end{sloppypar}[2]& $\kappa_{\mathrm{2-10keV}}$ \newline[3]&$\mathrm{log(M_{BH}/M_{\odot})}$ [4]&Comments [5]\\\hline

NGC 4395&0.001064&$0.000104\pm0.000026$&$0.00234\pm0.00047$&$22.5^{+13.0}_{-8.1}$&$5.56$\\
NGC 3227&0.003859&$0.00907\pm0.00021$&$0.1301\pm0.0021$&$14.35^{+0.57}_{-0.55}$&$7.63$\\
NGC 6814&0.006&$0.000970\pm0.000077$&$0.01954\pm0.00060$&$20.1^{+2.4}_{-2.0}$&$7.08$\\
NGC 3516&0.008836&$0.078\pm0.020$&$1.03\pm0.20$&$13.2^{+8.0}_{-4.7}$&$7.63$\\
NGC 4593&0.009&$0.071\pm0.013$&$0.467\pm0.059$&$6.6^{+2.5}_{-1.7}$&$6.73$\\
NGC 3873&0.01&$0.149\pm0.014$&$2.13\pm0.12$&$14.3^{+2.4}_{-2.0}$&$7.04$\\
NGC 7469&0.016&$0.163\pm0.015$&$3.38\pm0.10$&$20.8^{+2.8}_{-2.4}$&$6.88$\\
Mrk 79&0.022&$0.178\pm0.044$&$2.05\pm0.30$&$11.5^{+6.1}_{-3.6}$&$8.01$\\
Ark 564&0.025&$0.177\pm0.040$&$1.62\pm0.25$&$9.2^{+4.5}_{-2.9}$&$6.90$&NLS1\\
Mrk 335&0.026&$0.158\pm0.013$&$14.98\pm0.14$&$94.8^{+9.5}_{-8.1}$&$6.58$&NLS1\\
Mrk 279&0.03&$0.538\pm0.042$&$9.00\pm0.23$&$16.7^{+1.9}_{-1.6}$&$7.54$\\
Mrk 290&0.03&$0.186\pm0.020$&$2.28\pm0.12$&$12.3^{+2.1}_{-1.7}$&$8.05$\\
Ark 120&0.032&$0.659\pm0.044$&$10.35\pm0.33$&$15.7^{+1.7}_{-1.4}$&$8.27$\\
Mrk 509&0.034&$1.31\pm0.15$&$14.6\pm1.2$&$11.2^{+2.5}_{-2.0}$&$7.96$\\
ESO 141-G55&0.036&$0.504\pm0.042$&$25.69\pm0.25$&$51.0^{+5.1}_{-4.3}$&$8.85$\\
KUG 1031+398&0.042&$0.0376\pm0.0054$&$3.293\pm0.091$&$88^{+17}_{-13}$&$5.80$&NLS1\\
NGC 985&0.042&$0.366\pm0.034$&$5.59\pm0.15$&$15.3^{+2.0}_{-1.7}$&$8.05$\\
Mrk 506&0.043&$0.476\pm0.037$&$4.55\pm0.21$&$9.6^{+1.3}_{-1.1}$&$7.47*$\\
Fairall 9&0.047&$1.126\pm0.085$&$8.23\pm0.64$&$7.3^{+1.2}_{-1.0}$&$7.92$\\
3C 382&0.058&$3.91\pm0.32$&$38.0\pm1.7$&$9.7^{+1.3}_{-1.1}$&$8.97$&RL\\
PG 1011-040&0.058&$0.01068\pm0.00078$&$10.807\pm0.064$&$1012^{+86}_{-74}$&$7.19$&XRW\\
I Zw 1&0.061&$0.249\pm0.060$&$4.79\pm0.46$&$19.3^{+8.5}_{-5.2}$&$7.24$&NLS1\\
Ton S180&0.062&$0.527\pm0.033$&$49.26\pm0.68$&$93.4^{+7.7}_{-6.8}$&$7.09$&NLS1\\
II Zw 136&0.063&$0.49\pm0.16$&$13.34\pm0.94$&$27.1^{+16.0}_{-8.1}$&$7.80$\\
PG 1229+204&0.063&$2.23\pm0.33$&$17.2\pm1.6$&$7.7^{+2.2}_{-1.6}$&$7.93$\\
Ton 951&0.064&$0.232\pm0.017$&$16.71\pm0.17$&$71.9^{+6.7}_{-5.7}$&$7.76$\\
MR 2251-178&0.066&$2.583\pm0.016$&$56.83\pm0.29$&$22.00^{+0.25}_{-0.25}$&$6.90$\\
Mrk 205&0.071&$0.963\pm0.029$&$8.14\pm0.13$&$8.46^{+0.40}_{-0.37}$&$7.75$\\
Mrk 478&0.079&$0.381\pm0.028$&$27.47\pm0.89$&$72.1^{+8.2}_{-7.1}$&$7.33$&NLS1\\
PG 1211+143&0.081&$0.522\pm0.045$&$69.5\pm2.0$&$133^{+17}_{-14}$&$7.37$&NLS1\\
Mrk 1383&0.086&$2.05\pm0.15$&$68.7\pm1.1$&$33.5^{+3.3}_{-2.8}$&$8.92$\\
PG 0804+761&0.1&$2.08\pm0.20$&$59.4\pm1.1$&$28.5^{+3.7}_{-3.0}$&$8.21$\\
PG 1415+451&0.114&$0.366\pm0.089$&$13.95\pm0.54$&$38.1^{+14.0}_{-8.6}$&$7.80$\\
Mrk 876&0.129&$1.38\pm0.45$&$113.0\pm2.8$&$82^{+43}_{-22}$&$8.95$\\
PG 1626+554&0.133&$1.44\pm0.36$&$27.8\pm2.1$&$19.3^{+8.4}_{-5.0}$&$8.37$\\
PG 0026+129&0.142&$4.40\pm0.30$&$147.9\pm2.5$&$33.6^{+3.1}_{-2.7}$&$7.42$\\
PG 1114+445&0.144&$1.30\pm0.11$&$18.1\pm1.3$&$14.0^{+2.4}_{-2.0}$&$8.41$\\
PG 1352+183&0.152&$1.242\pm0.093$&$17.92\pm0.64$&$14.4^{+1.7}_{-1.5}$&$8.30$\\
PG 1115+407&0.154&$0.810\pm0.067$&$23.86\pm0.74$&$29.5^{+3.7}_{-3.1}$&$7.50$&NLS1\\
PG 0052+251&0.155&$4.66\pm0.33$&$76.4\pm2.5$&$16.4^{+1.8}_{-1.6}$&$8.74$\\
PG 1307+085&0.155&$1.26\pm0.28$&$77.0\pm2.0$&$61^{+20}_{-12}$&$8.52$\\
3C 273&0.158&$83.0\pm9.9$&$1220\pm95$&$14.7^{+3.3}_{-2.6}$&$8.54$&RL\\
PG 1402+261&0.164&$1.75\pm0.17$&$122.7\pm1.9$&$70.0^{+8.5}_{-7.0}$&$7.76$&NLS1\\
PG 1048+342&0.167&$1.11\pm0.23$&$28.1\pm1.7$&$25.3^{+8.6}_{-5.6}$&$8.24$\\
PG 1322+659&0.168&$0.608\pm0.046$&$32.46\pm0.47$&$53.4^{+5.2}_{-4.5}$&$8.08$\\
PG 2349-014&0.174&$5.13\pm0.36$&$115.1\pm2.1$&$22.4^{+2.2}_{-1.9}$&$9.26$\\
PG 1116+215&0.176&$4.94\pm0.65$&$176.2\pm5.5$&$35.7^{+6.7}_{-5.1}$&$8.70$\\
4C +34.47&0.206&$8.43\pm0.56$&$76.0\pm3.6$&$9.01^{+1.10}_{-0.97}$&$8.50$&RL\\
PG 0947+396&0.206&$2.27\pm0.44$&$52.3\pm2.6$&$23.1^{+7.0}_{-4.7}$&$8.53$\\
PG 0953+414&0.243&$3.92\pm0.80$&$263.7\pm4.2$&$67^{+19}_{-12}$&$8.49$\\
PG 1444+407&0.267&$1.55\pm0.13$&$99.54\pm0.86$&$64.1^{+6.3}_{-5.4}$&$8.54$\\
PG 1100+772&0.311&$6.61\pm0.64$&$216.1\pm3.4$&$32.7^{+4.1}_{-3.4}$&$9.11$&RL\\
PG 1216+069&0.331&$7.69\pm0.73$&$184.9\pm3.4$&$24.1^{+3.0}_{-2.5}$&$8.95$\\
PG 1512+370&0.371&$9.3\pm2.0$&$173\pm12$&$18.6^{+6.7}_{-4.3}$&$9.17$&RL\\

\hline
\end{tabular}
\caption{Bolometric corrections for all AGN in the sample.  [1] \210keV luminosity from \textsc{xspec} fit.  May differ slightly from the values in the source literature due to slight variations in the best fit parameters. [2] Bolometric accretion luminosity is the total luminosity in the energy range $0.001-250\mathrm{keV}
$. [3] \210keV bolometric correction.  Errors on bolometric corrections are `max/min'; i.e. they represent the range of values achievable when the input parameters are varied within their error limits. [4] Logarithm of Black Hole mass $\mathrm{M_{BH}}$ in units of solar mass. *The mass estimate for Mrk 506 was taken from a study with systematically higher masses than in other studies used here, so the mass estimate for Mrk 506 has been scaled down appropriately. [5] ``NLS1'' denotes a Narrow Line Seyfert 1 nucleus, ``RL'' denotes a radio loud AGN and ``XRW'' denotes an X-ray weak AGN. }
\label{table:bolcorstable}
\end{table*}

\subsection{Checking SED shapes against the Literature}

\begin{figure}
   \includegraphics[width=8.5cm]{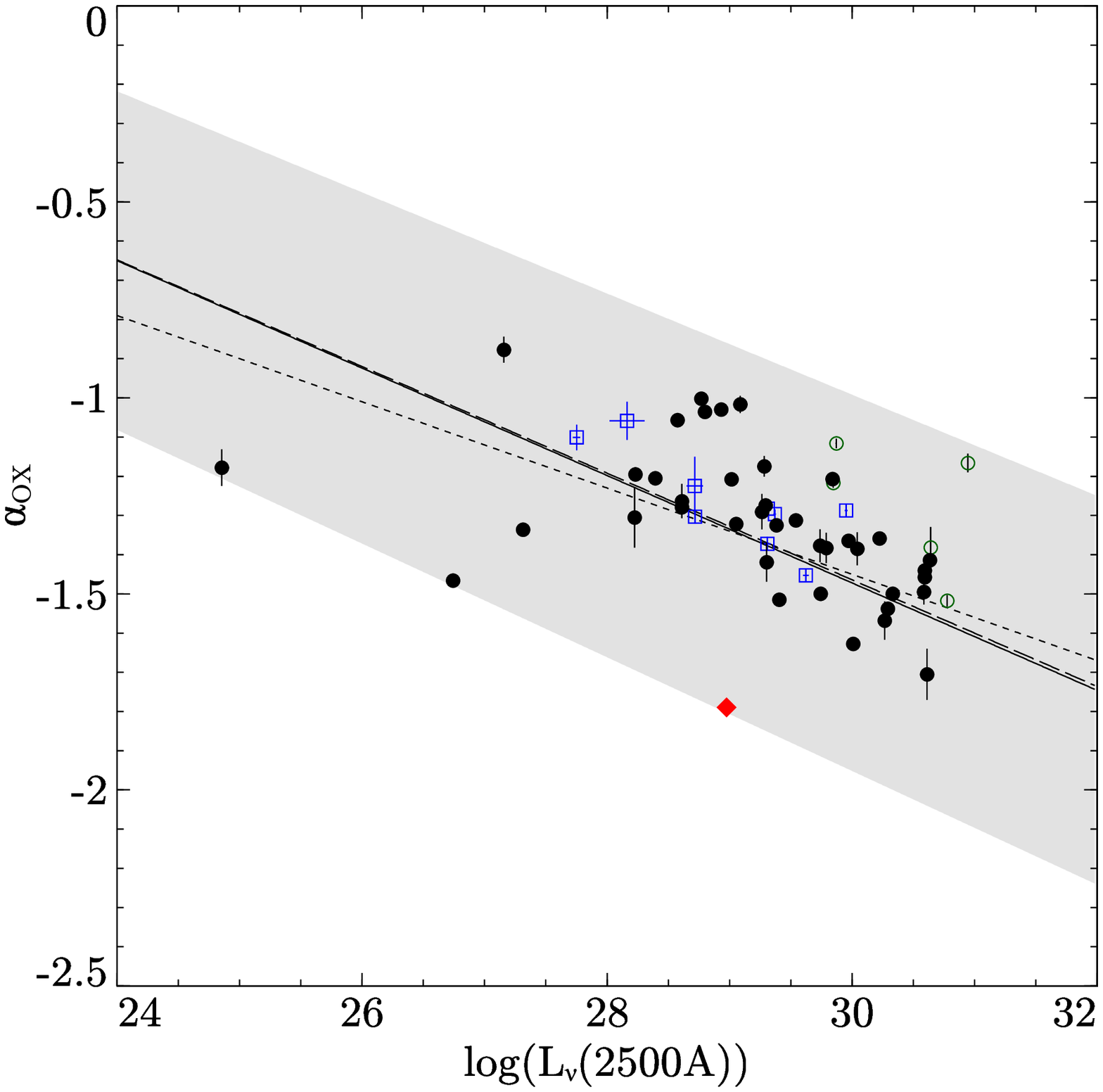}
   \caption{Plot of $\mathrm{\alpha_{OX}}$ against $2500\mathrm{\AA}$ monochromatic luminosity for the AGN in our sample. Key as in Fig. \ref{bcvslumallagn}.  The solid line represents the best fit to the \protect\cite{2006AJ....131.2826S} sample, with the dashed line representing the best fit from the study of \protect\cite{2005AJ....130..387S}.  The dotted line represents the best fit from \protect\cite{2003AJ....125.2876V}, provided for reference and comparison with Fig. 11 of \protect\cite{2005AJ....130..387S}.  Grey shaded area gives the spread in the Steffen et al. best fit.}
\label{alphaOXvsUVlum}
\end{figure}

The anti-correlation between $\mathrm{\alpha_{OX}}$ and $\mathrm{2500\AA}$ luminosity discussed in Section \ref{Intro} usefully provides a constraint on the physical processes at work, as it tells us that the fraction of power in the accretion disc corona (emitted in the X-rays) decreases with increasing UV luminosity. \cite{2005AJ....130..387S} used 228 optically selected AGN across a range of redshifts ($0.01<z<6.3$) to provide a robust characterisation of this relation.  They confirm that the $\mathrm{\alpha_{OX}-L_{\nu}(2500\AA})$ relation represents an underlying physical correlation rather than an accumulation of systematics. \cite{2006AJ....131.2826S} extend their work to a larger range in luminosities by including fainter AGN (from the Extended \emph{Chandra} Deep Field--South) and more luminous AGN (from the Bright Quasar Survey).  This provides the most recent and robust determination of the relation, spanning the largest space in the luminosity--redshift plane.
We compare the values of $\mathrm{\alpha_{OX}}$ generated from our SEDs to the best fit of \cite{2006AJ....131.2826S} in Fig. \ref{alphaOXvsUVlum}.

It can be seen that the values for $\mathrm{\alpha_{OX}}$ obtained lie within the limits of the \cite{2006AJ....131.2826S} fit, implying sensible reconstruction of the SED shapes.

\section{Discussion of Systematics}

Fig. \ref{bcvslumallagn} shows that the bolometric corrections obtained do not convincingly follow the forms suggested by \cite{2004MNRAS.351..169M} and \cite{2006astro.ph..5678H}; the reasons for this spread could be diverse, ranging from selection effects to intrinsic properties of the AGN SED. In this section we outline a number of systematic effects which may affect the bolometric correction for individual AGN.

\subsection{Intrinsic Exctinction}

The shape of the BBB in the optical--UV region of AGN spectra is notoriously prone to effects of reddening.  Without properly correcting for de-reddening effects, it is impossible to know the intrinsic emission of AGN in these wavebands and therefore it becomes difficult to constrain energy output from AGN.  Galactic extinction has been investigated by numerous authors and values of the Galactic reddening in the standard form $E(B-V)$ are readily available for the AGN in our sample, along with standard Galactic extinction curves.  All the observations presented in the \cite{2004ApJ...615..135S} and \cite{2005ApJ...619...41S} samples have already been de-reddened for Galactic extinction according to the extinction curve of \cite{1989ApJ...345..245C} with values for Galactic extinction provided by \cite{1998ApJ...500..525S}.

\begin{figure}
    \includegraphics[width=4cm]{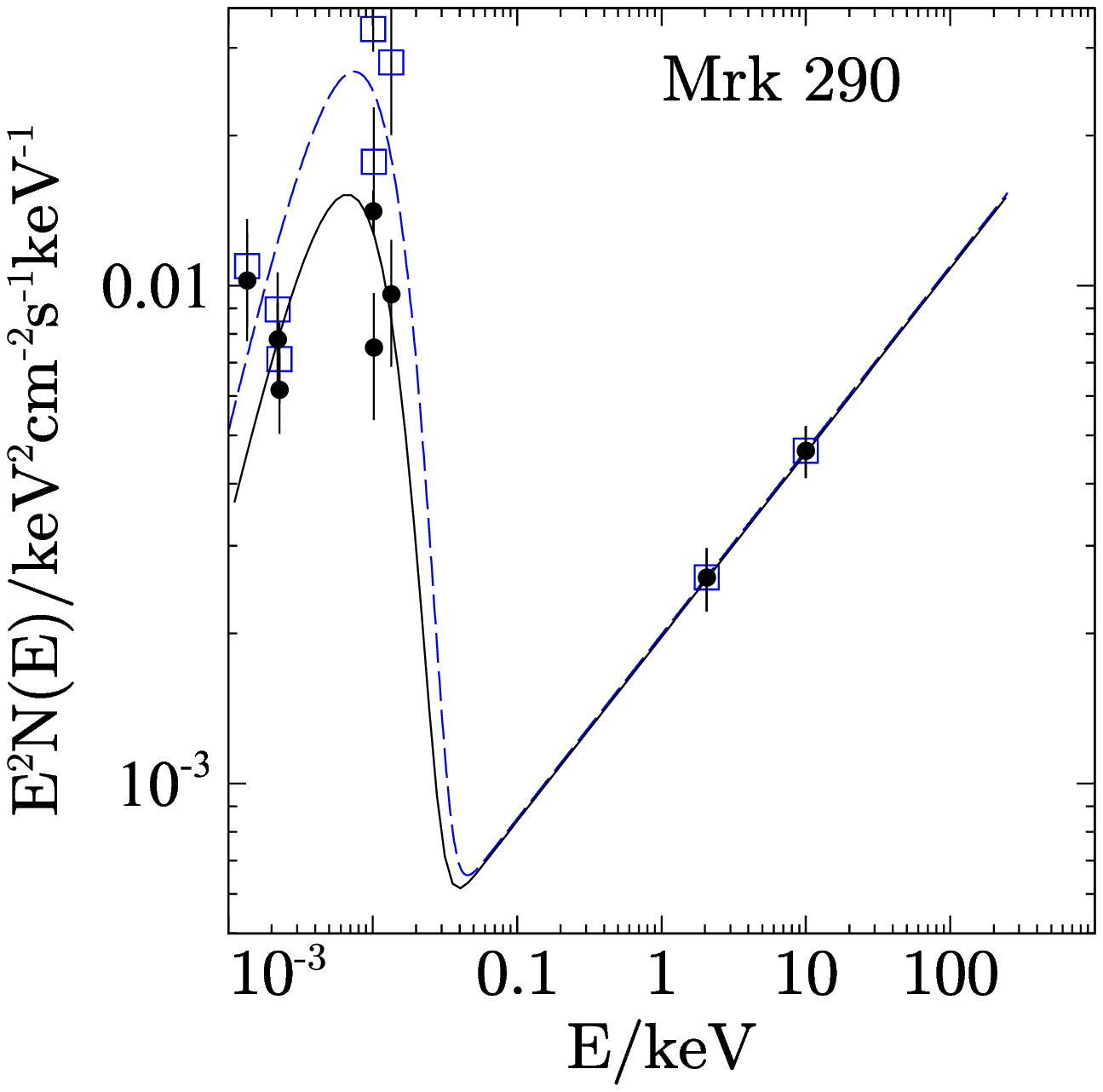}
    \includegraphics[width=4cm]{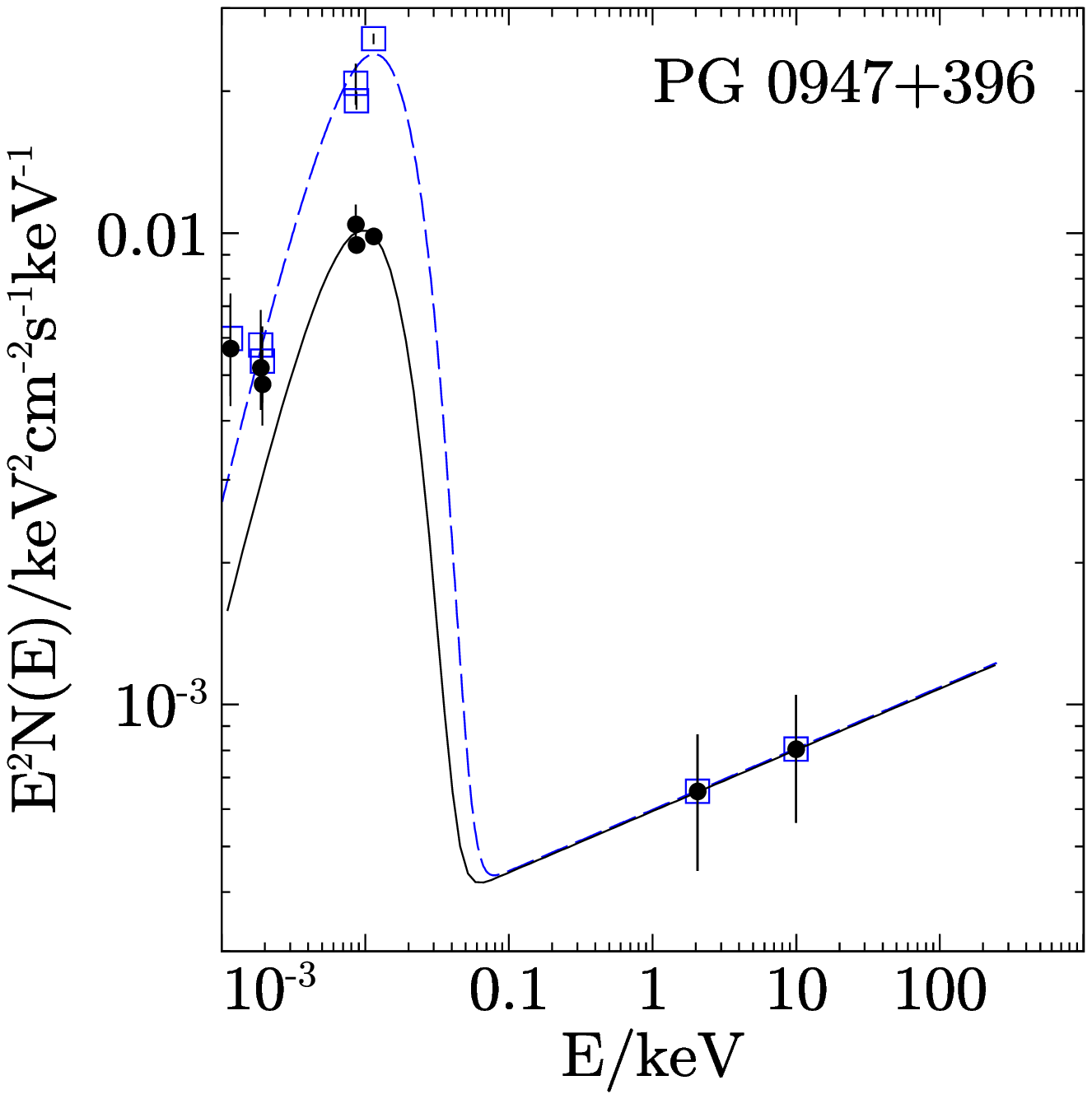}
    \includegraphics[width=4cm]{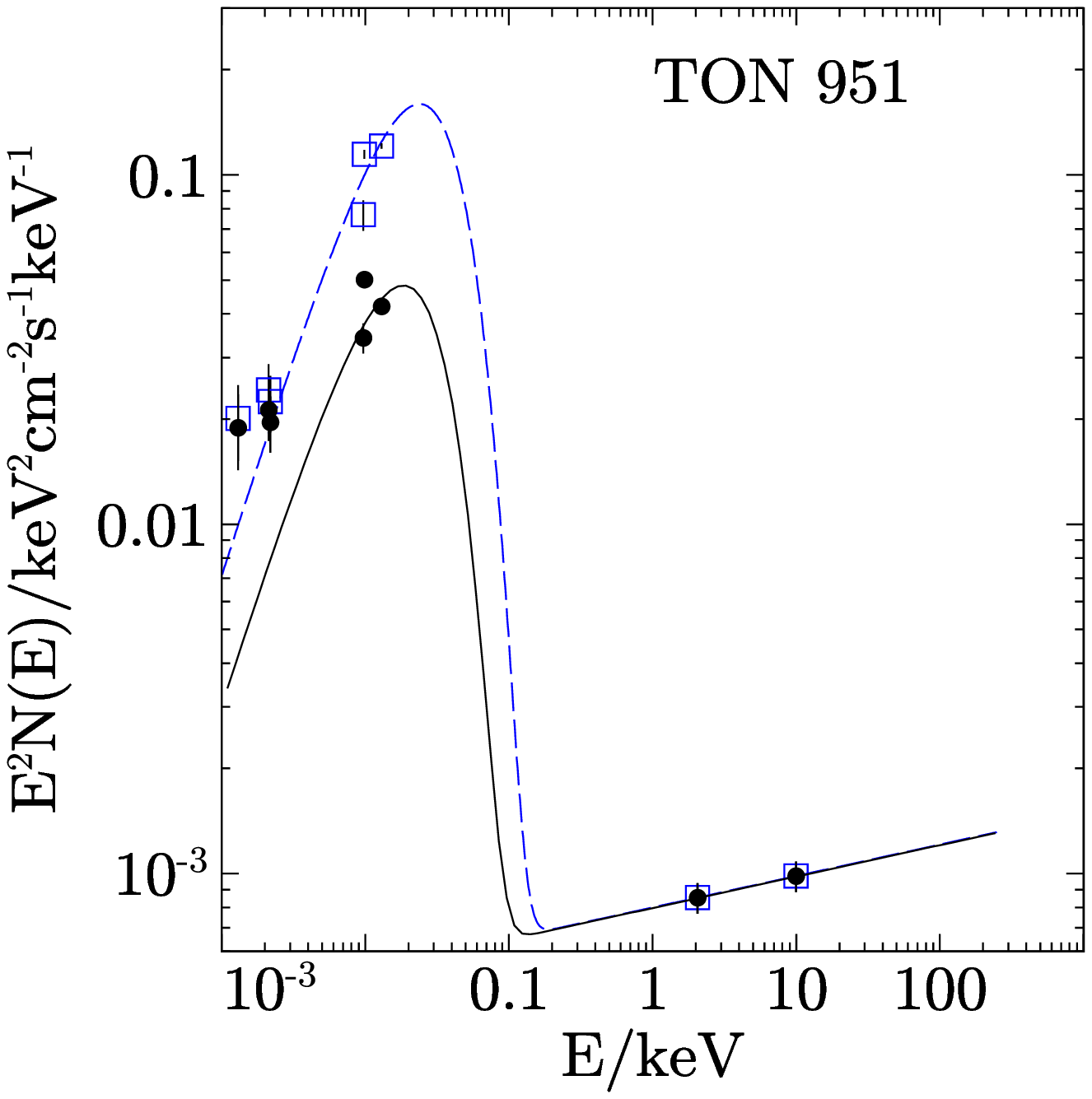}
    \includegraphics[width=4cm]{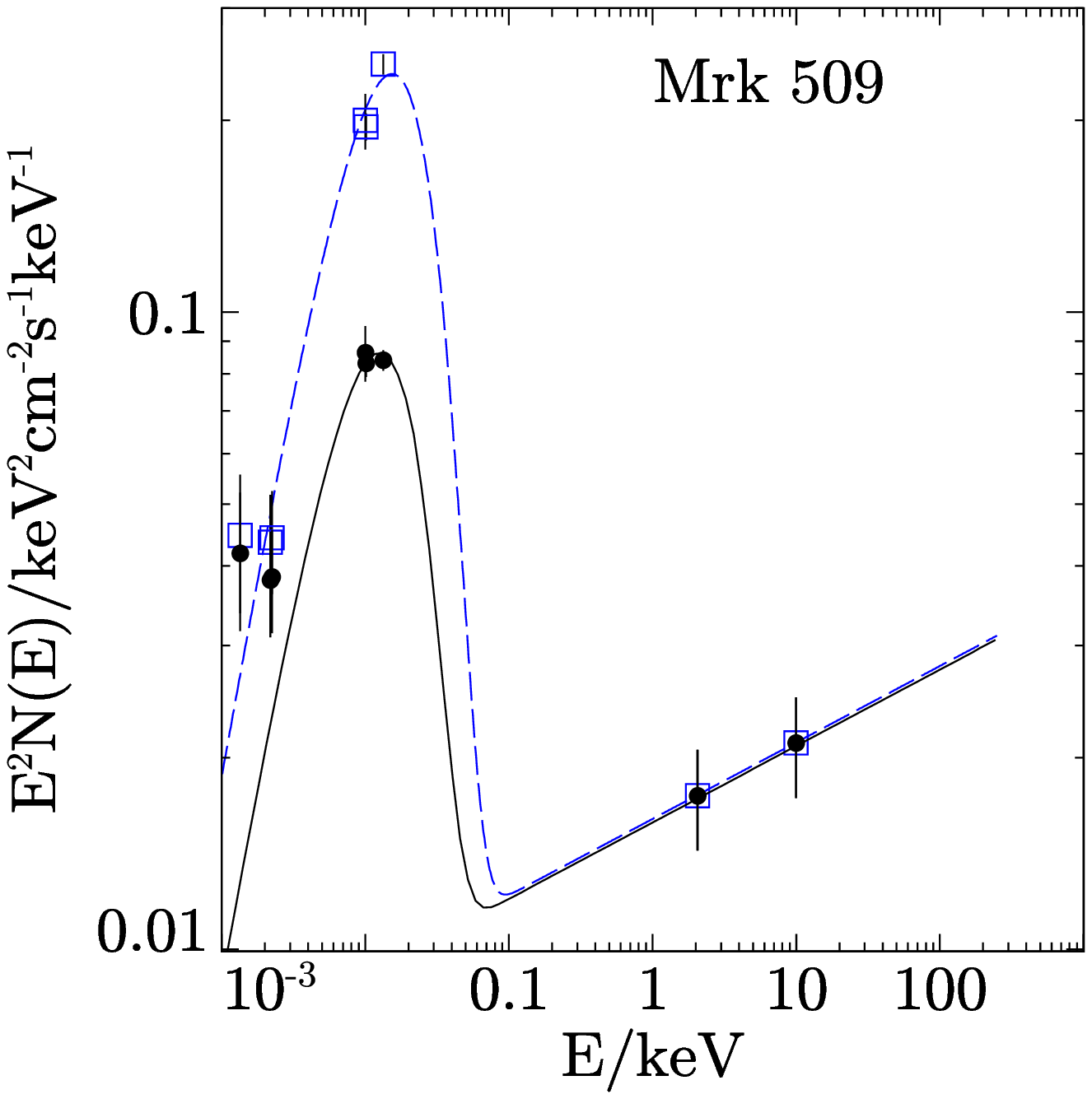}
    \caption{Comparison of a sample of SEDs with and without de-reddening due to intrinsic extinction (using the SMC extinction curve of \protect\citealt{1992ApJ...395..130P}) assuming a reddening of $E(B-V)=0.055$. Filled black circles and black solid lines are the data and model fit to the pre-dereddened data respectively, whereas empty blue squares and blue dashed lines are the data and model SED fits to the de-reddened data points.}
\label{SMCreddeningSEDs}
\end{figure}

However the process of accounting for intrinsic extinction due to the AGN itself is more problematic, and such reddening can vary significantly between different AGN. There remains much debate over the shape of the wavelength dependence of the extinction.  Forms which have been proposed included Small Magellanic Cloud-type (SMC) extinction, Milky Way-type exction (both derived by \citealt{1992ApJ...395..130P}), or a curve which is flatter in the far UV \citep{2004ApJ...616..147G}.  \cite{2004AJ....128.1112H} find that an SMC-type extinction curve best accounts for the reddening in a sample of Sloan Digital Sky Survey (SDSS) quasars, in preference to Mikly Way-type or \cite{2004ApJ...616..147G}-type reddening.  They identify that $\sim 91$ per cent of the sample has reddening $E(B-V)<0.055$ (taking into account the effects of quasars being `extincted out' due to unusually high reddenings).  To obtain a relatively conservative estimate of the effects of SMC-type reddening on the bolometric correction, we apply this reddening curve along with $E(B-V)=0.055$ to the 12 AGN from our sample coincident with the \cite{2005ApJ...619...41S} (for which the optical--UV coverage is more complete than the rest of our objects).  We find that applying such de-reddening produces an increase in $\kappa_{\mathrm{2-10keV}}$ of a factor of $\sim 2.0$.  A sample of the SEDs generated including de-reddening corrections fo SMC-type extinction are reproduced in Fig. \ref{SMCreddeningSEDs}.

In as yet unpublished work, \cite{2005AAS...20714901G} used the 17 AGN in the \cite{2005ApJ...619...41S} sample to produce a revised mean extinction curve for AGN, by comparing each AGN SED with three template `low-reddening' objects: 3C 273, PG 1100+772 and PG 0953+414.  These templates cover three different classes of AGN (namely radio loud, lobe dominated and radio quiet) thereby hoping to average out at least some of the effects of intrinsic object-to-object spectral variations in determining the average optical--UV intrinsic extinction curve.  In the process of normalising their extinction curves, they obtain $E(B-V)$ values for each of the AGN in their sample.

\begin{figure}
    \includegraphics[width=4cm]{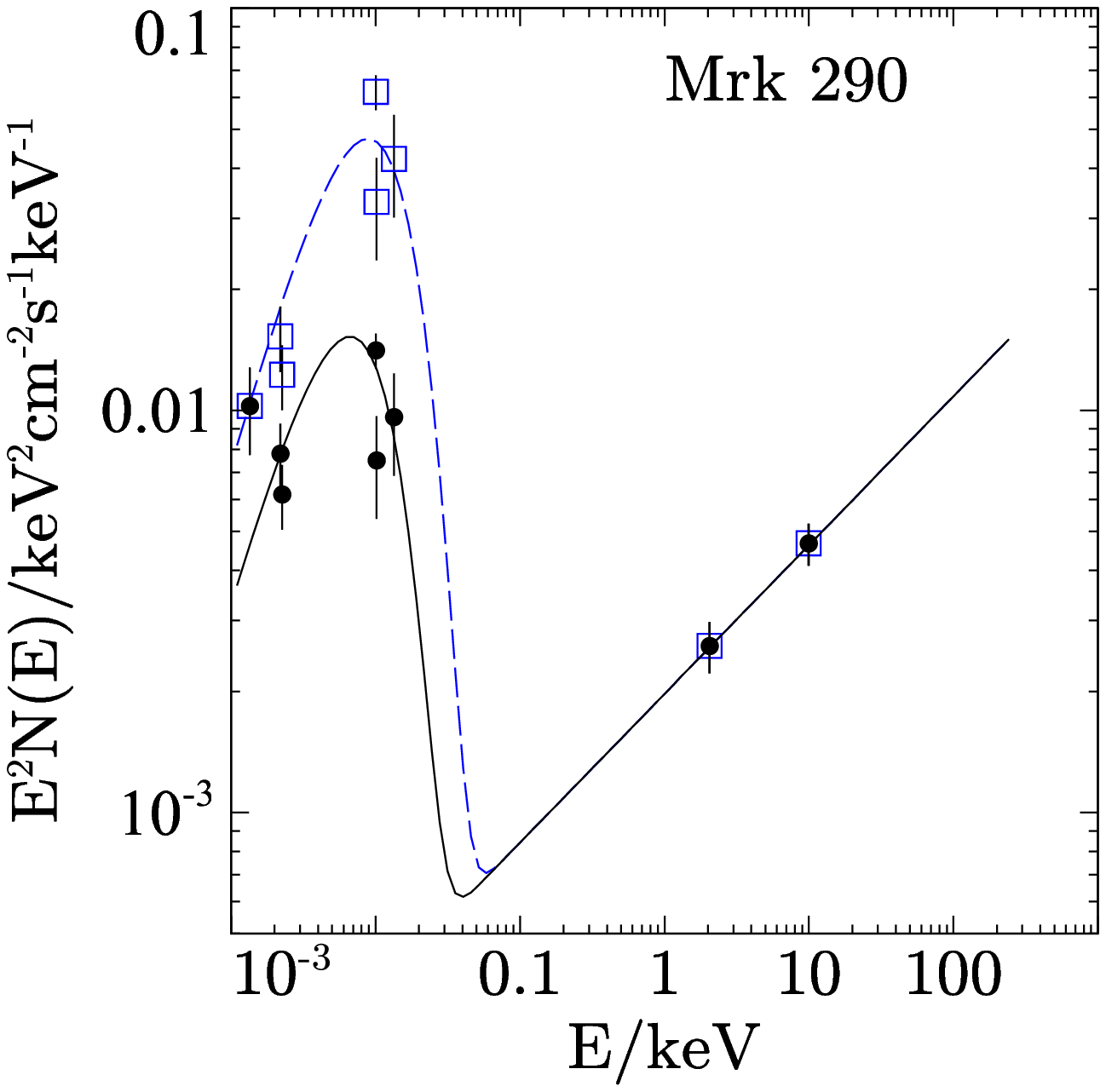}
    \includegraphics[width=4cm]{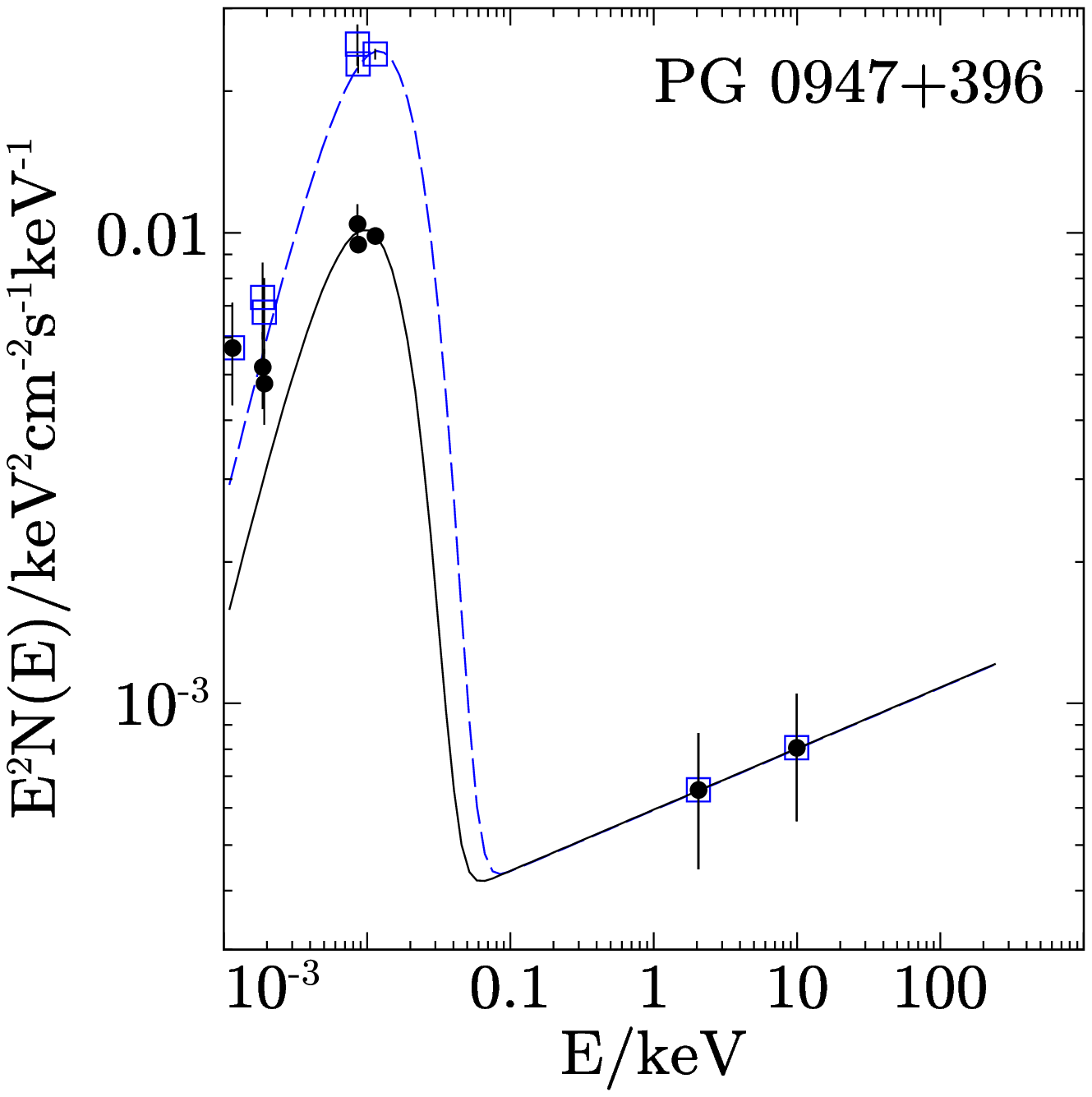}
    \includegraphics[width=4cm]{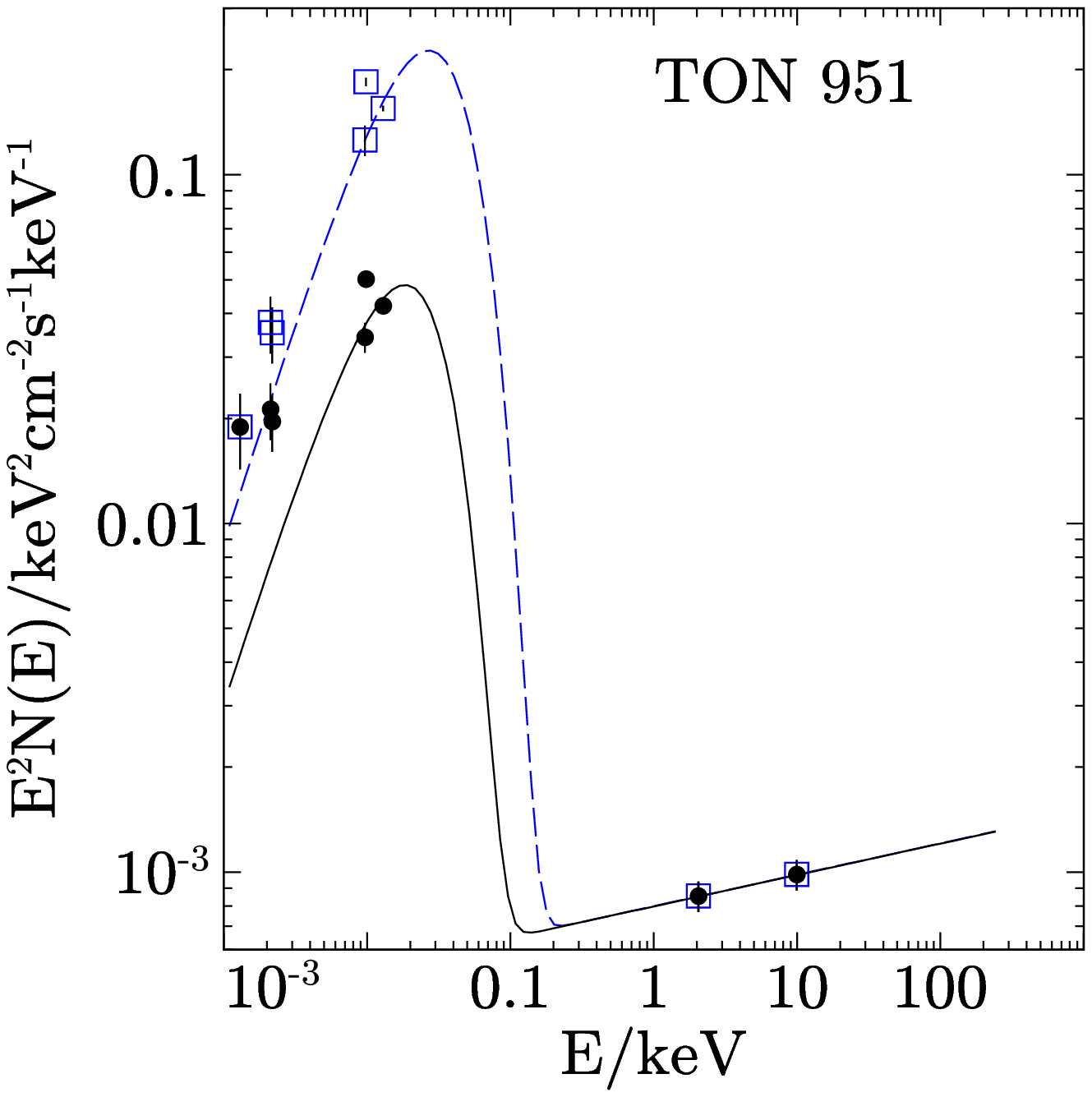}
    \includegraphics[width=4cm]{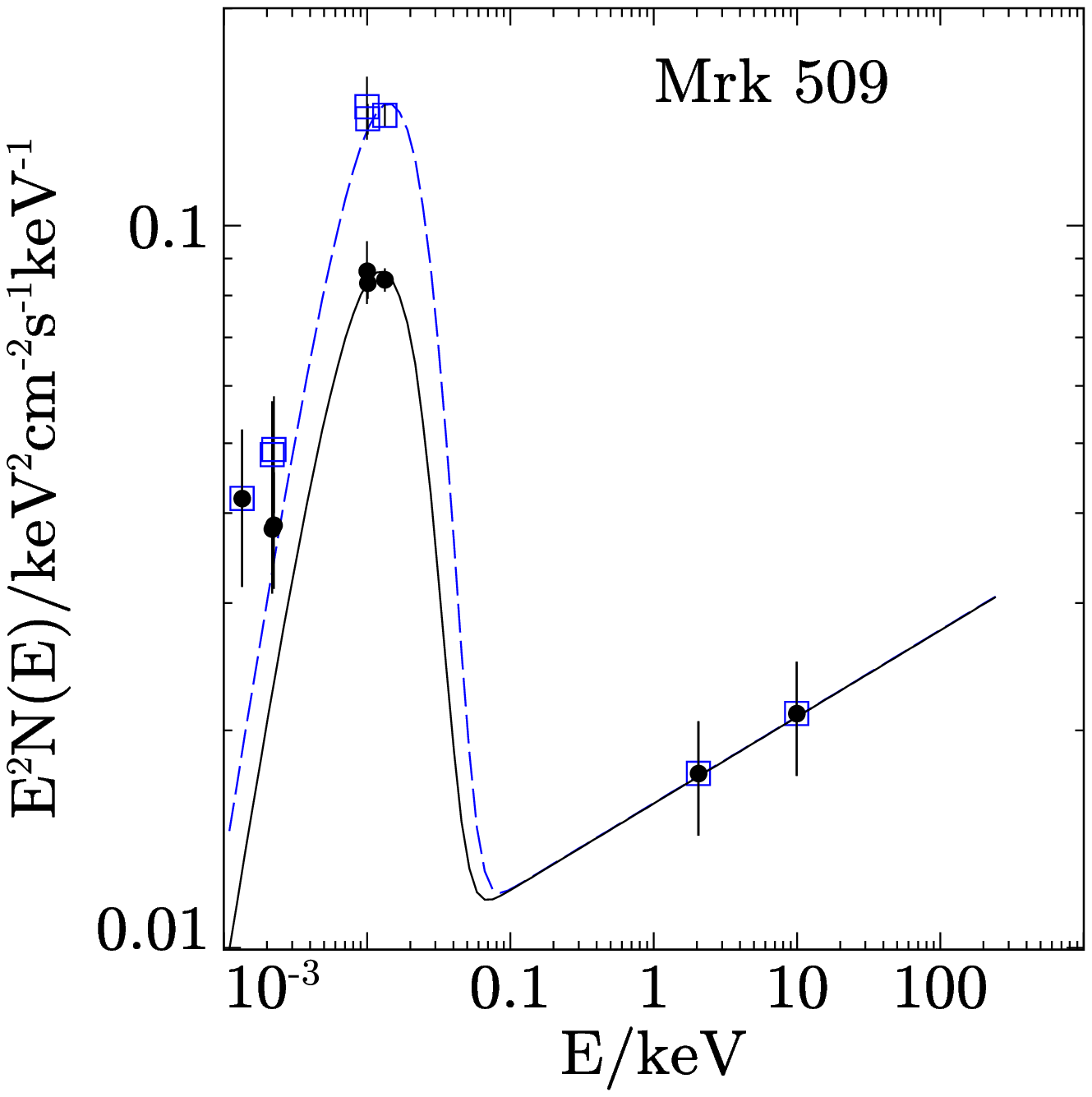}
    \caption{Comparison of a sample of SEDs with and without de-reddening due to intrinsic extinction (using the extinction curve of \protect\citealt{2005AAS...20714901G}). Key as in Fig. \ref{SMCreddeningSEDs}.}
\label{reddeningSEDs}
\end{figure}

\begin{figure}
    \includegraphics[width=9cm]{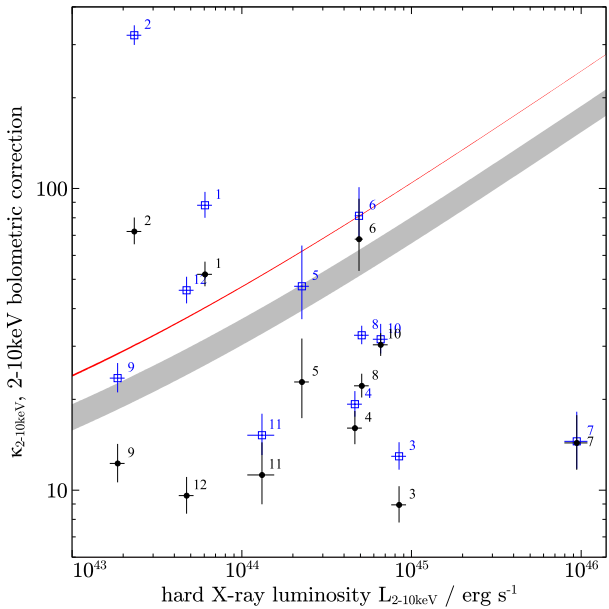}
    \caption{Comparison of bolometric corrections with and without de-reddening due to intrinsic extinction (using the extinction curve of \citealt{2005AAS...20714901G}). The filled black circles represent bolometric corrections before de-reddening, the open blue squares give the corresponding corrections after de-reddening. Other conventions as in Fig. \ref{bcvslumallagn}. The AGN in this subsample: 1 - PG 1322+659, 2 - TON 951, 3 - 4C +34.47, 4 - PG 0052+251, 5 - PG 0947+396, 6 - PG 0953+414, 7 - 3C 273, 8 - PG 2349-014, 9 - Mrk 290, 10 - PG 1100+772, 11 - Mrk 509, 12 - Mrk 506.}
\label{reddening}
\end{figure}

We also used their mean extinction curve along with the values of $E(B-V)$ specific to each AGN from \cite{2005AAS...20714901G} to construct de-reddened SEDs for the 12 AGN coincident with the \cite{2005ApJ...619...41S} sample. A sample of these are plotted in Fig. \ref{reddeningSEDs}, with the resulting bolometric corrections given in Fig. \ref{reddening}.  An increase in bolometric correction from de-reddened SEDs is again clearly visible; the average increase is a factor of $\sim 1.9$, very similar to that for SMC reddening.  The use of individual values of $E(B-V)$ for each AGN also hints at an increase in reddening at lower luminosities.  In summary, we estimate that the maximum effect of intrinsic reddening is to increase $\kappa_{\mathrm{2-10keV}}$ by a factor $\sim 2$.


An overarching study of reddening mechanisms in AGN, covering the luminosity range from Quasars to Seyferts would be extremely illuminating and allow considerable progress to be made in understanding the AGN SED.  Nonetheless, it is clear from our results that the effect of optical--UV reddening will be to increase the X-ray bolometric correction, which will have a well defined effect on the SMBH mass density estimates from the XRB.  If de-reddening uniformly increases bolometric corrections as seen here, then we require a greater average mass--to--radiation efficiency, $\eta$ to satisfy the Soltan argument.  This would imply more rapidly spinning black holes and reduces the relative importance of radiatively inefficient modes of accretion in the X-ray background.

\subsection{Narrow Line Seyfert 1 Nuclei, Radio Loud AGN and X-ray Weak AGN}

Narrow Line Seyfert 1 (NLS1) nuclei are well known to exhibit spectral complexity at X-ray energies; this manifests as sharp spectral drops or gradual curvature in the spectrum \citep{2006MNRAS.368..479G}.  The common paradigm for NLS1 nuclei assumes a central black hole of lower mass than that typical for other classes of AGN, with either ionized reflection or partial covering giving rise to the characteristic narrow lines in the optical spectrum. \cite{2006MNRAS.368..479G} also discusses the likelihood that NLS1s could represent the nucleus in either a low or normal flux state, and that NLS1 nuclei could switch between the two on typical X-ray variability timescales for AGN.  We note in Fig. \ref{bcvslumallagn} that a large number of bolometric corrections for NLS1 nuclei are seen to lie systematically above the form suggested by \cite{2004MNRAS.351..169M}.  NLS1 nuclei seem to represent an interesting class of object which have a larger spread in bolometric correction than `normal' AGN, with a possible marginal skew towards higher bolometric corrections.

We have hitherto not made any consideration for the radio properties of these AGN, despite the known links between their X-ray and radio emission \citep{1993ApJ...410...29S}.  Strongly radio loud objects potentially present complications when determining the X-ray emission directly associated with accretion, as pointed out by \cite{2005AJ....130..387S} who exclude this class of objects from the determination of the $\mathrm{\alpha_{OX}-L_{\nu}(2500\AA)}$ relation.  This complication arises because the observed X-ray flux may contain combined emission from both accretion and jet processes (\citealt{1997ApJS..109..103D}); since we wish to calculate intrinsic accretion luminosities using the bolometric corrections, it is necessary to account for the energy output of the jet or exclude these objects altogether.  Accounting for the X-ray flux due to jet emission is problematic, as it is unclear what fraction of X-ray emission in radio loud AGN is from jets.  Fortunately, we are able to gain an estimate of the importance of this effect with one of our radio loud AGN, 3C 273 thanks to the recent publication of XMM observations taken at a historic jet minimum by \cite{2006A&A...451L...1T}.  Using their \210keV flux and photon index instead of ASCA data, we obtain a bolometric correction of $20.4^{+2.3}_{-2.0}$ compared to the value $14.7^{+3.3}_{-2.6}$ obtained using the ASCA flux.  A larger bolometric correction is in better concordance with the expected value from \cite{2004MNRAS.351..169M}, but still falls significantly short (the expected bolometric correction at this luminosity is between 150-200).  This may imply some significant jet contribution still remaining in the \citealt{2006A&A...451L...1T} flux, or the low bolometric correction for 3C 273 may simply be another example of the large spread we see in the rest of our results.

The curious object PG 1011-040 has a bolometric correction of $\sim1000$, the highest observed in our sample.  This AGN is identified by \cite{2001ApJ...546..795G} as a member of the `X-ray weak' AGN class, which exhibit particularly low ratios of soft X-ray flux to optical flux.  Our inclusion of multiwavelength data confirms this with the high bolometric correction.  These authors discuss the possibility of these unusually low X-ray fluxes being due to either intrisic X-ray weakness or a high level of X-ray absorption.  They note that PG 1011-040 displays no spectral evidence of significant X-ray absorption or variability, implying that the high bolometric correction we obtain is probably typical for the object and does not represent a snapshot of an unusual state.  They also point out that its low $\mathrm{H}\beta$ linewidth would place it in the NLS1 subclass.  

In general, a significant proportion of AGN may be X-ray weak, and a study by \cite{2000ApJ...528..637B} identifies $\sim 11$ per cent of the optically selected QSO sample of \cite{1992ApJS...80..109B} (the Bright Quasar Survey) as falling into this category.  If a similar figure holds for the general AGN population, the presence of a substantial proportion of AGN with bolometric corrections of order $\sim 1000$ could significantly affect the SMBH density calculated via the Soltan argument.  

\begin{figure*}
   \includegraphics[width=10cm]{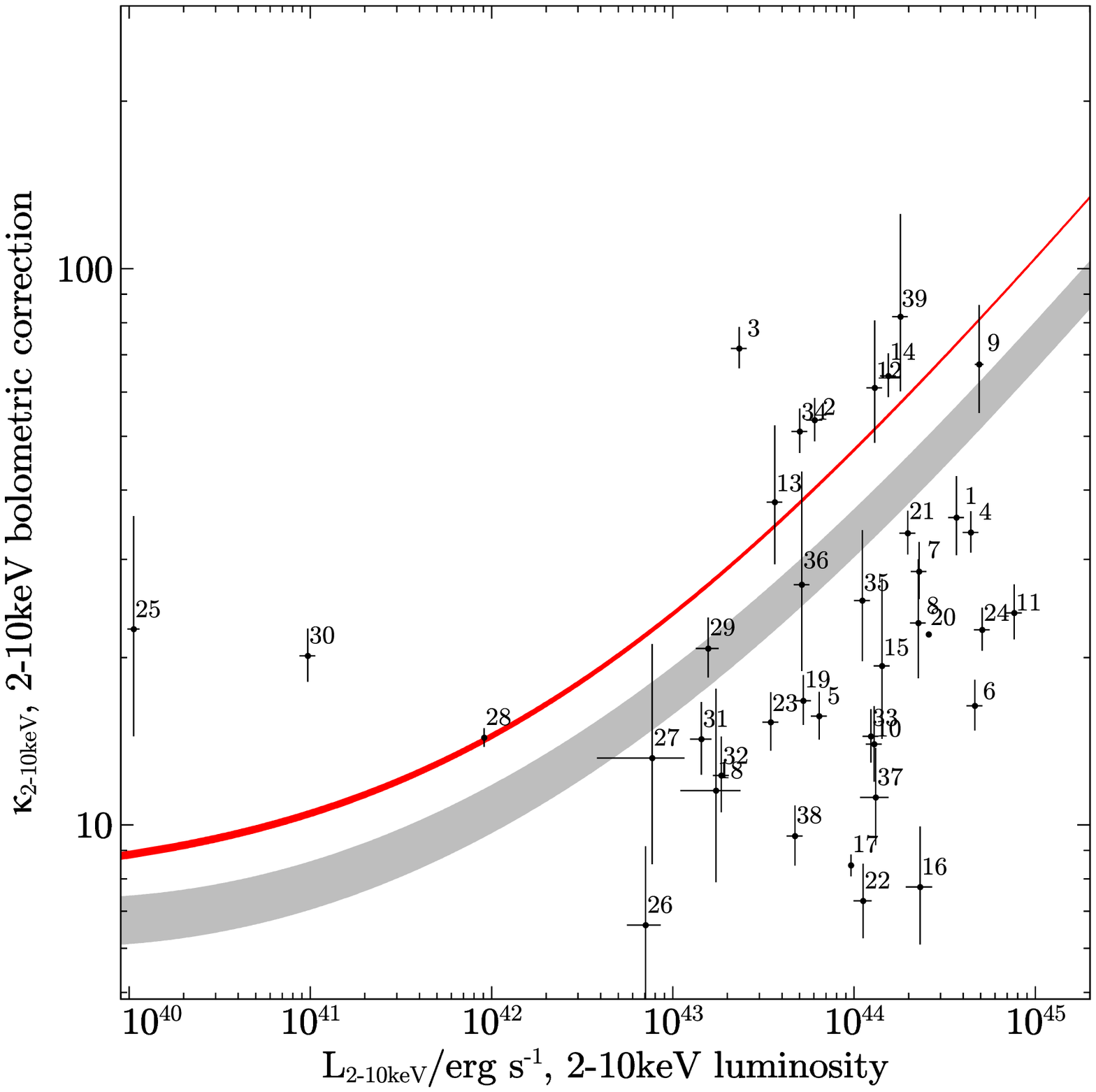}
   \includegraphics[width=10cm]{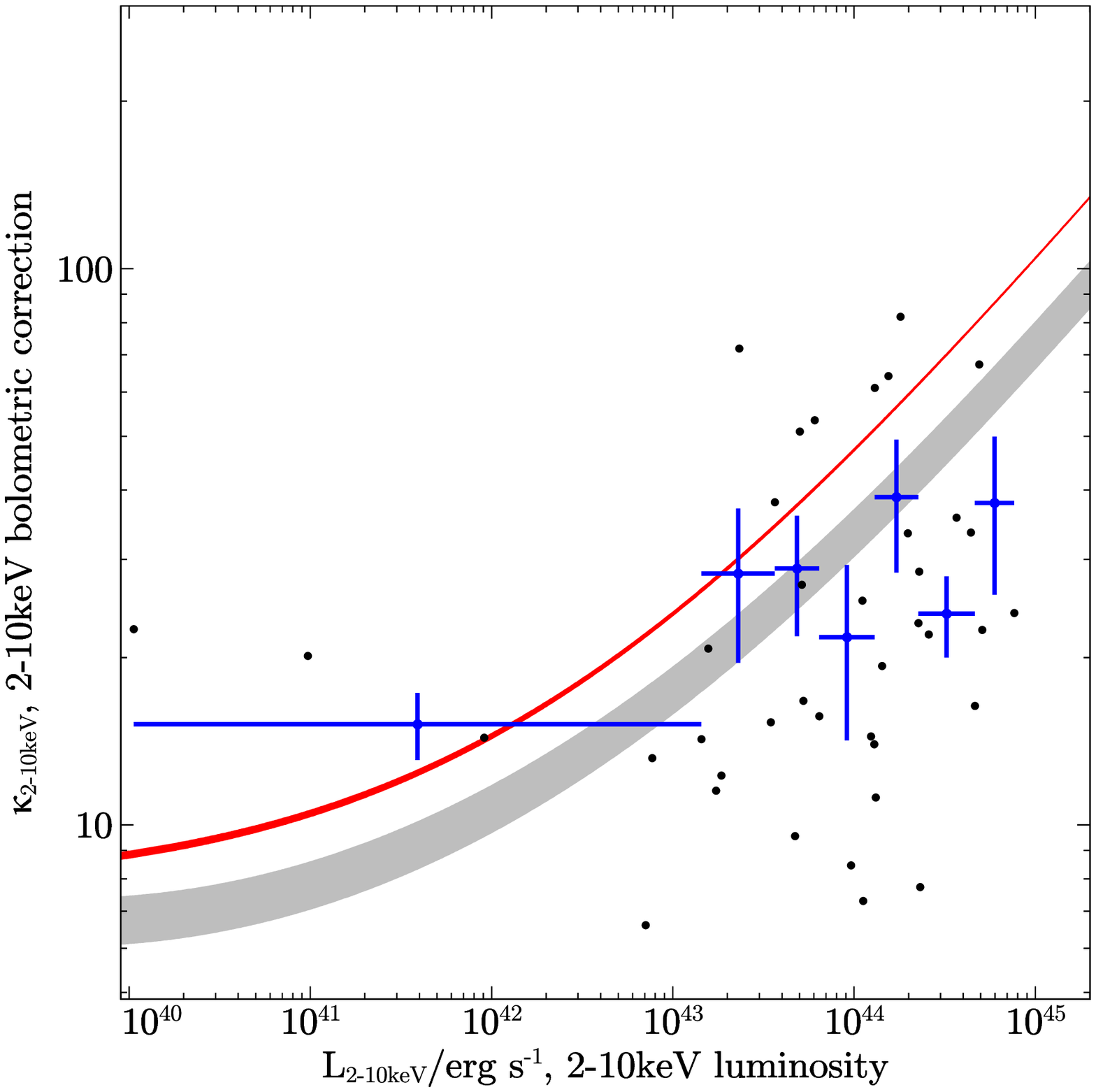}
   \caption{Hard X-ray Bolometric Correction against \210keV luminosity, with Radio Loud, NLS1 and X-ray weak AGN removed.  Lower panel: binned bolometric corrections, six AGN per bin.  Other conventions as in Figure \ref{bcvslumallagn}. The AGN in the subsample: 1 - PG 1116+215, 2 - PG 1322+659, 3 - Ton 951, 4 - PG 0026+129, 5 - Ark 120, 6 - PG 0052+251, 7 - PG 0804+761, 8 - PG 0947+396, 9 - PG 0953+414, 10 - PG 1114+445, 11 - PG 1216+069, 12 - PG 1307+085, 13 - PG 1415+451, 14 - PG 1444+407, 15 - PG 1626+554, 16 - PG 1229+204, 17 - Mrk 205, 18 - Mrk 79, 19 - Mrk 279, 20 - MR 2251-178, 21 - Mrk 1383, 22 - Fairall 9, 23 - NGC 985, 24 - PG 2349-014, 25 - NGC 4395, 26 - NGC 4593, 27 - NGC 3516, 28- NGC 3227*, 29 - NGC 7469, 30 - NGC 6814, 31 - NGC 3873, 32 - Mrk 290, 33 - PG 1352+183, 34 - ESO 141-G55, 35 - PG 1048+342, 36 - II Zw 136, 37 - Mrk 509, 38 - Mrk 506, 39 - Mrk 876. * See footnote for this object in Fig. \ref{bcvslumallagn}.}
\label{bcvslumallagnblacklistremoved}
\end{figure*}

Radio loud and X-ray weak AGN are known to exhibit deviations from the $\mathrm{\alpha_{OX}}-\mathrm{L_{\nu}(2500\AA)}$ relation.  NLS1 nuclei may have unusual bolometric corrections as a result of their spectral complexity.   It is evident from Fig. \ref{bcvslumallagn} that these objects also represent many of those which deviate significantly from the expected dependences (\citealt{2004MNRAS.351..169M} and \citealt{2006astro.ph..5678H}).  We therefore present a plot of bolometric corrections against luminosity with these classes of objects removed in Fig. \ref{bcvslumallagnblacklistremoved}. These sources are also highlighted in the $\mathrm{\alpha_{OX}-L_{\nu}(2500\AA)}$ plot in Fig. \ref{alphaOXvsUVlum}.  We now find a weak increase in bolometric correction with luminosity with these objects removed, but with far more deviation than the forms suggested in the literature.

\subsection{UV and Optical variability: Simultaneity of SED data}

The presence of AGN variability in the X-ray and UV bands and the links between variability in the two bands are well documented (e.g. \citealt{2001ApJ...561..162S}).  For the AGN in this sample, SED data points have been taken from different publications in the literature, calculated from observations which are not in general contemporaneous.  While we might expect delays of a few days between observations to be insignificant and in accordance with the time delays expected between X-rays and optical wavelengths, the fact that the observations used to construct the SEDs are often separated by many years implies that SED data points in different bands could be providing snapshots of the SED in epochs with particularly incongruous spectral shapes.  This could partially account for some of the poorer model fits for some AGN in the sample.  The lack of simultaneous SED data could therefore represent a significant obstacle to estimating bolometric corrections accurately. 

\begin{figure}
   \includegraphics[width=9cm]{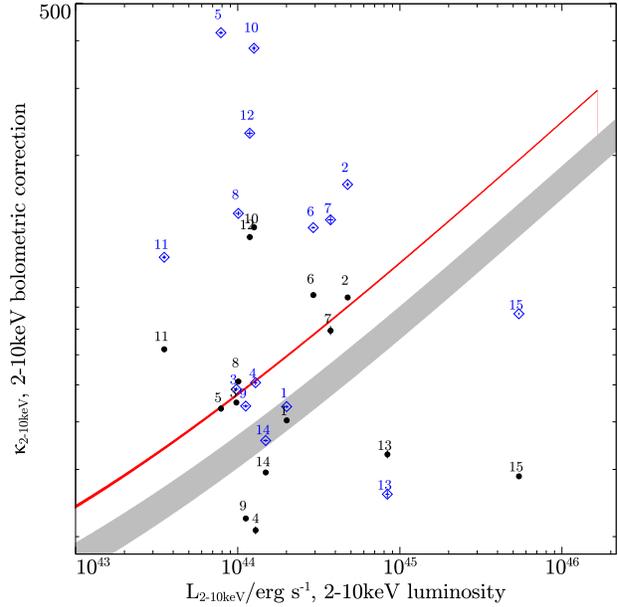}
   \caption{Bolometric corrections generated from simultaneous SEDs (open blue diamonds - using XMM Optical Monitor optical--UV data and XMM X-ray fluxes as given in \protect\citealt{2006MNRAS.366..953B}) and non-simultaneous SEDs (filled black circles - \emph{FUSE} UV points, miscellaneous optical, XMM X-ray fluxes as given in \protect\citealt{2006MNRAS.366..953B}). Other conventions as in Fig. \ref{bcvslumallagn}.  The AGN in the subsample: 1 - PG 0947+396, 2 - PG 0953+414, 3 - PG 1048+342, 4 - PG 1114+445, 5 - PG 1115+407, 6 - PG 1116+215, 7 - PG 1216+069, 8 - PG 1322+659, 9 - PG 1352+183, 10 - PG 1402+261, 11 - PG 1415+451, 12 - PG 1444+407, 13 - PG 1512+370, 14 - PG 1626+554, 15 - 3C 273.}
\label{fig:simultaneity}
\end{figure}

\cite{2006MNRAS.366..953B} present simultaneous optical--UV--X-ray points for 23 AGN, using XMM Optical Monitor (XMM-OM) data for optical and UV points. Fourteen are common to our sample, allowing us to gauge the importance of simultaneity in determining accurate SEDs. These authors generate SED fits to the data using a more detailed Comptonized disc model than the \textsc{diskpn} model used in our study, but for the purposes of calculating bolometric corrections, our simple model should be sufficient.  Indeed, fitting the data points in \cite{2006MNRAS.366..953B} to the \textsc{diskpn} and power law combination yields better fits than our non-contemporaneous datasets for these AGN.  The bolometric corrections calculated from these fits are presented in Fig. \ref{fig:simultaneity} with the bolometric corrections calculated from non-contemporaneous datasets presented for comparison.

\begin{figure}
   \includegraphics[width=4cm]{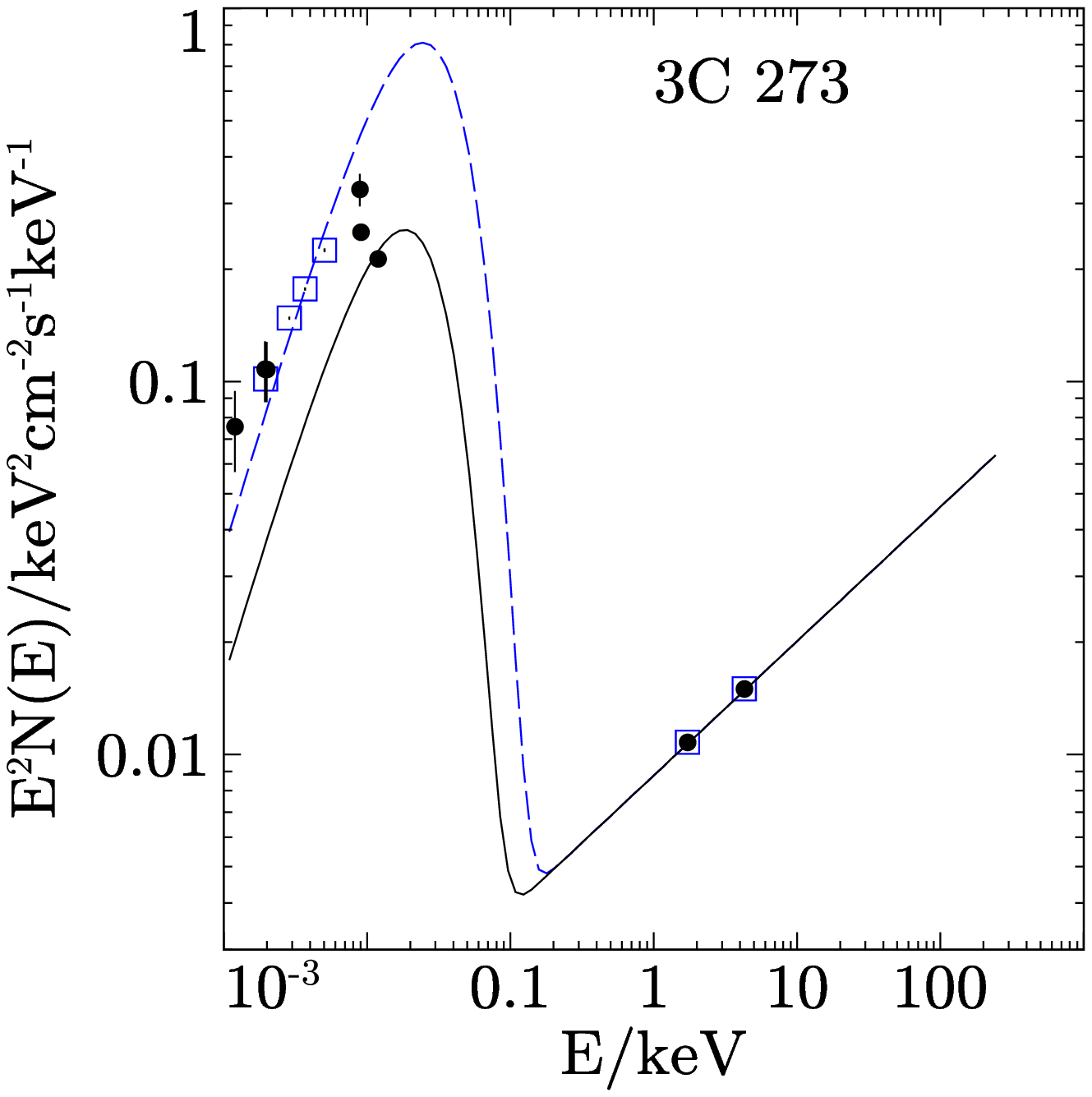}
   \includegraphics[width=4cm]{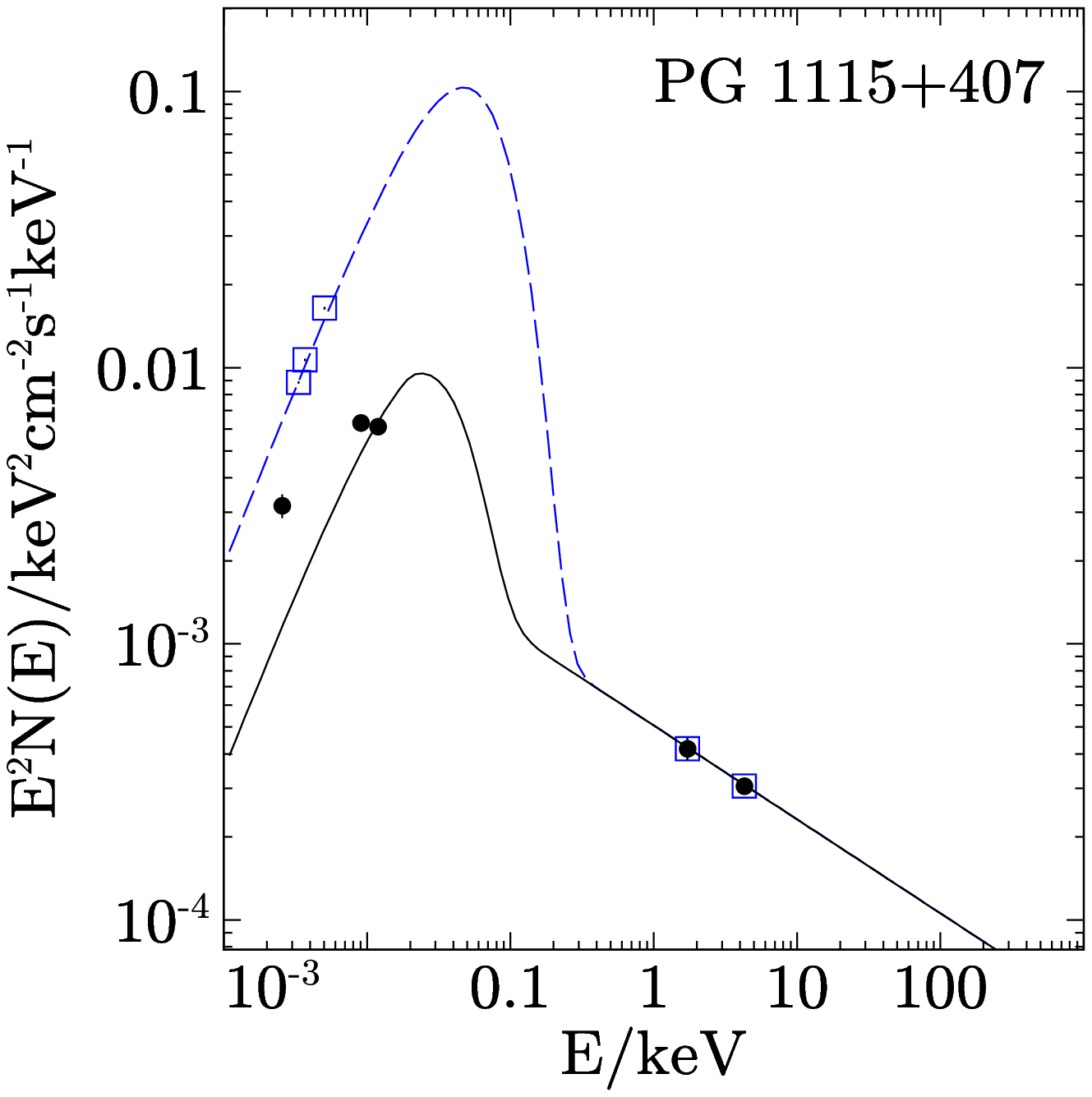}
   \includegraphics[width=4cm]{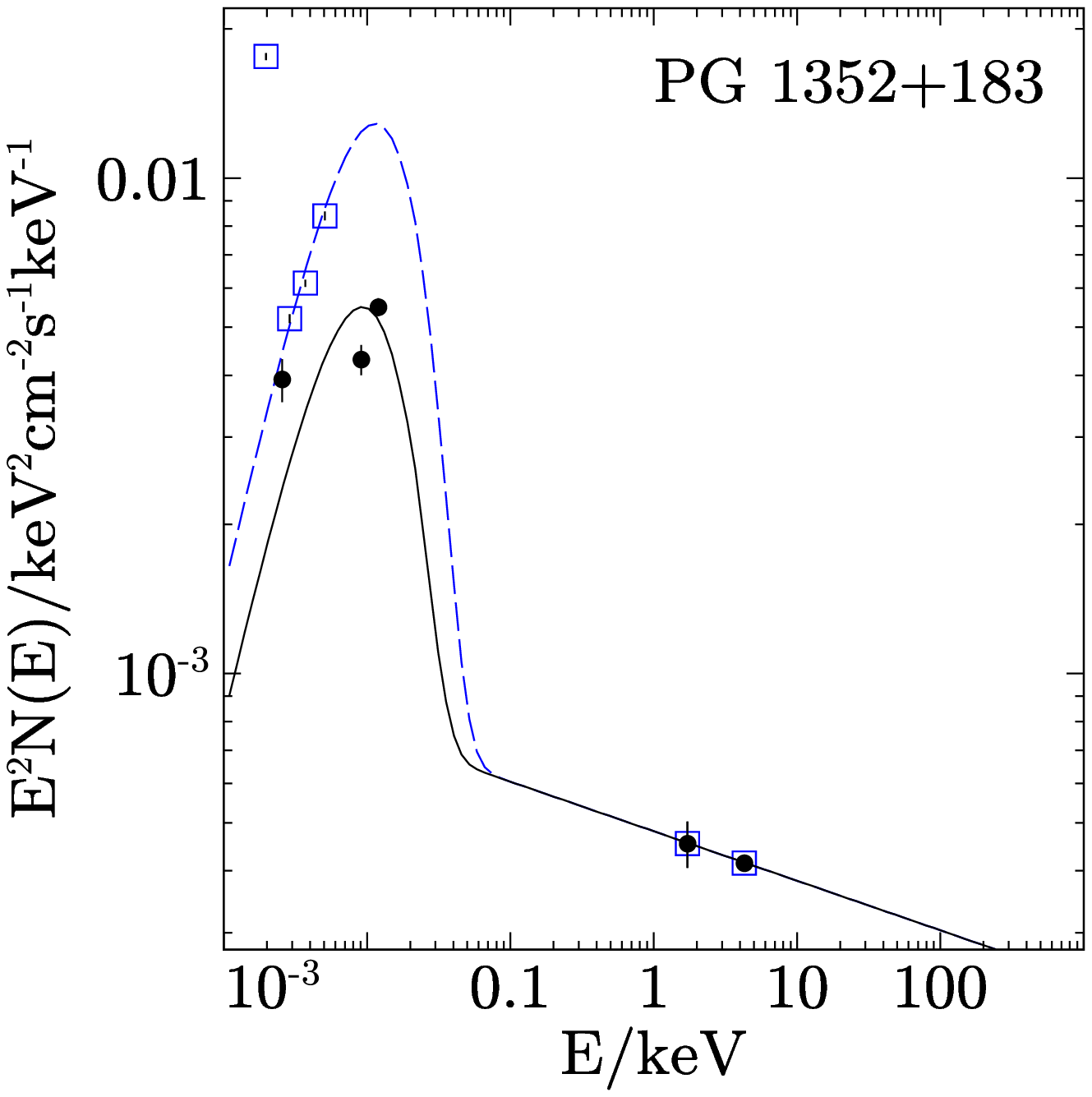}
   \includegraphics[width=4cm]{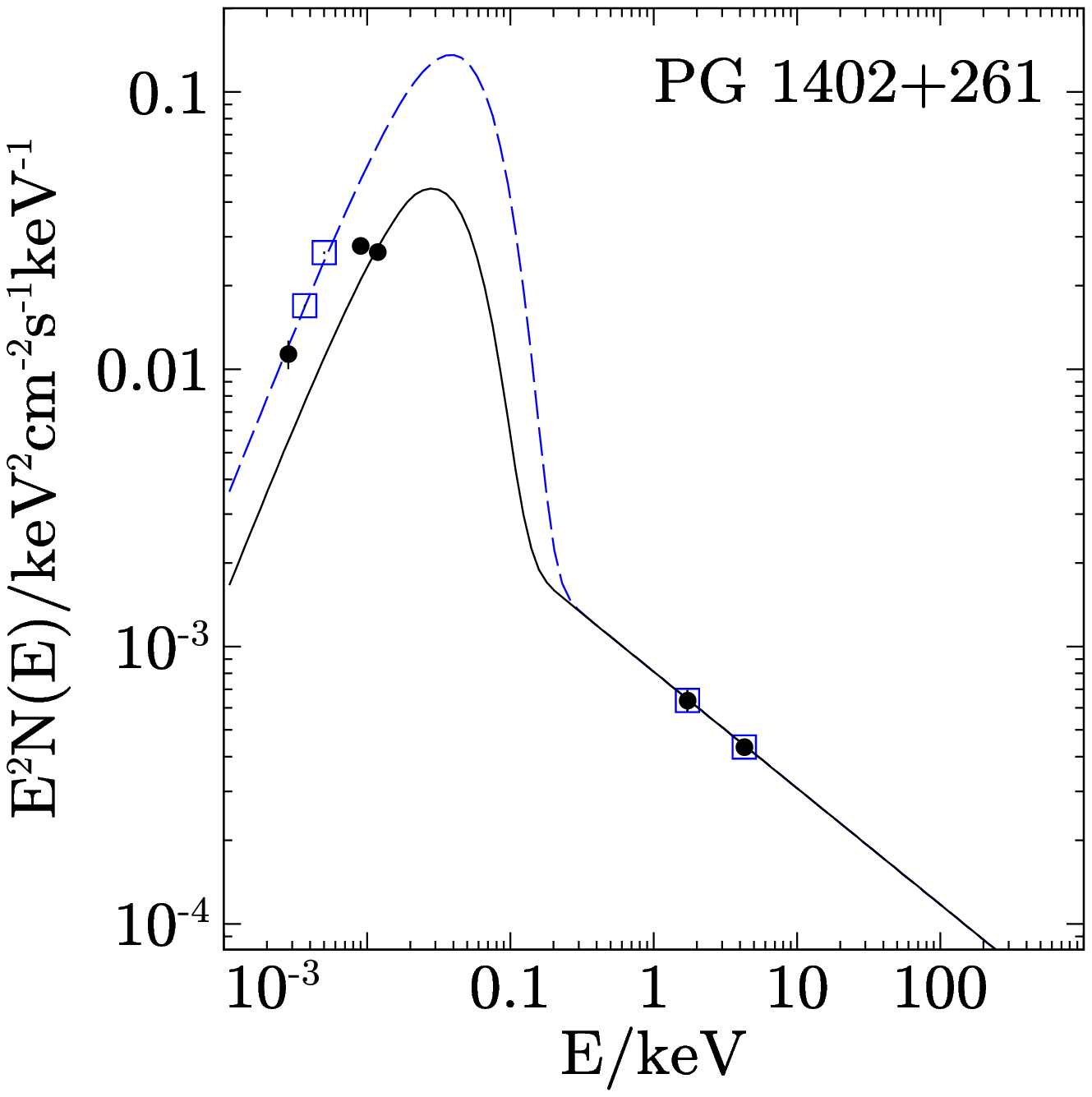}
   \caption{Comparing a sample of AGN SEDs from simultaneous data (XMM X-ray and XMM-OM optical and UV points) and non-simultaneous data (\emph{FUSE} UV, miscellaneous optical, XMM X-ray fluxes from \protect\citealt{2006MNRAS.366..953B}).  Filled cricles represent non-simultaneous data points, open squares represent simultaneous points.}
\label{fig:simultaneityseds}
\end{figure}

We see from Fig. \ref{fig:simultaneity} that the bolometric corrections from simultaneous SEDs trace out similar relative positions to each other as for non-simultaneous SEDs, but are generally higher than for non simultaneous SEDs.  Inspection of the SEDs (Fig. \ref{fig:simultaneityseds}) reveals that this is primarily due to the XMM-OM points (between wavelengths $\mathrm{2120\AA < \lambda < 5483\AA}$) missing the peak of the BBB emission in the UV; this allows the model to peak higher in flux than when constrained by the \emph{FUSE} UV points.  As discussed previously, the dip in the \emph{FUSE} UV points at the peak of the BBB could be due to intrinsic reddening, but we do not explore this issue further here.  It is important to note from this that we can only recording the true accretion luminosity if we use simultaneous SEDs; this motivates a further study with a larger simultaneous sample.

\subsection{X-ray Variability}

For those AGN for which multiple ASCA observations were available, the average and standard deviation X-ray fluxes and photon indicies were employed which feeds directly into the determination of errors on the bolometric correction, thereby taking into account some element of X-ray variability.  In other cases, where multiple observations from different satellites or different publications in the literature were available, it is not robust to take averages due to potentially differing systematics.  In these cases we compare the ratios of \210keV fluxes from the different satellites/publications in table \ref{table:variabilitytable}.  For the 12 AGN where observations from both ASCA and XMM were available, we find that there is no identifiable systematic offset between fluxes from the two satellites, with XMM fluxes exceeding ASCA fluxes $58$ per cent of the time.

We see that a typical value for the ratio `maximal flux to minimal flux' is $\sim1.5$ enabling us to estimate the effect of this variation on the bolometric correction.  For example, considering 3C 273 with max/min flux ratio 2.19, the bolometric corrections generated from the different X-ray datasets are $14.7^{+3.3}_{-2.6}$ (ASCA), $28.786^{+0.085}_{-0.085}$ (XMM -- \citealt{2006MNRAS.366..953B}) and $20.4^{+2.3}_{-2.0}$ (XMM -- \citealt{2006A&A...451L...1T}). This gives a ratio of maximum bolometric correction to minimum bolometric correction of $\sim2$.  It is also worth bearing in mind that some AGN are known to have significantly higher variability than the subsample highlighted here, a case in point being NGC 4395 \citep{2005MNRAS.356..524V} which has been observed to rapidly vary in X-rays by factors of 10.  There is therefore an obvious tendency for variability to increase the spread in bolometric corrections, and this adds to the uncertainties in their determination.

\begin{table*}
\begin{tabular}{|l|l|l|r|r}
\hline
AGN&\begin{sloppypar}Available X-ray observations [1]\end{sloppypar}&Observation with maximal&Maximal $2-10\mathrm{keV}$flux&Maximal flux/Minimal flux \\
 & & $2-10\mathrm{keV}$ flux & $(10^{-12} \mathrm{erg/s/cm^2})$& \\\hline

3C 273&ASCA (T),&ASCA (T)&98.91&2.19\\
 & XMM (B), & & & \\
 & XMM (\protect{\citealt{2006A&A...451L...1T}}) & & & \\
I Zw 1&ASCA (T),&XMM (P)	&4.83&2.86\\
 & XMM (P) & & & \\
Mrk 478	&ASCA (T),	&XMM (P)	&3.18	&1.59\\
 & XMM (P) & & & \\
PG 0947+396	&XMM (B),&XMM (P)	&1.73	&1.18\\
 &  XMM (P) & & & \\
PG 0953+414	&ASCA (T), &XMM (P) 	&2.80	&1.26\\
 &  XMM (B), & & & \\
 &  XMM (P) & & & \\
PG 1048+342	&XMM (B),	&XMM (B)	&1.37	&1.15\\
 &  XMM (P) & & & \\
PG 1100+772	&ASCA (T),	&XMM (P)	&3.07	&1.67\\
 & XMM (P) & & & \\
PG 1114+445	&ASCA (T),	&XMM (B)	&2.43	&1.09\\
 & XMM (B), & & & \\
 & XMM (P) & & & \\
PG 1115+407	&XMM (B),	&XMM (P)	&1.21	&1.10\\
 &  XMM (P) & & & \\
PG 1116+215	&ASCA (T),	&ASCA (T)	&3.70	&1.29\\
 & XMM (B) & & & \\
PG 1216+069	&ASCA (T),	&ASCA (T)	&1.54	&2.20\\
 &  XMM (B), & & & \\
 &  XMM (P) & & & \\
PG 1322+659	&ASCA (T), &XMM (P)	&1.12	&1.66\\
 &  XMM (B), & & & \\
 &  XMM (P) &&&\\
PG 1352+183	& XMM (B),	&XMM (P)	&1.88	&1.14\\
 & XMM (P) &&&\\
PG 1402+261	&ASCA (T),	&ASCA (T)	&1.95	&1.45\\
 &  XMM (B) &&&\\
PG 1415+451	&XMM (B), &XMM (P)	&1.06	&1.07\\
 & XMM (P) &&&\\
PG 1444+407	&ASCA (T), 	&ASCA (T)	&0.59	&1.41\\
 &  XMM (B) &&&\\
PG 1512+370	&XMM (B),	&XMM (P)	&1.87	&1.11\\
 & XMM (P) &&&\\
PG 1626+554	&XMM (B),	&XMM (B)	&3.12	&1.03\\
 & XMM (P) &&&\\
TON 951		&ASCA (T), 	&XMM (P)	&3.93	&2.33\\
 &  XMM (P) &&&\\
\hline
\end{tabular}
\caption{Hard X-ray variability for AGN for which multiple satellite observations were available. [1] ASCA (T) denotes ASCA observations from the Tartarus database, XMM (B) denotes XMM observations as presented in \protect\cite{2006MNRAS.366..953B} and XMM (P) denotes XMM observations as presented in \protect\cite{2005A&A...432...15P}. }
\label{table:variabilitytable}
\end{table*}

\subsection{Effects of Variations in SMBH mass estimates}

The normalisation of the disk component for each of the AGN in the sample is constrained using the mass and luminosity distance to the source.  The bolometric corrections presented so far do not reflect uncertainties in SMBH mass estimates; accounting for such uncertainties is difficult due to the diverse methods used to obtain these estimates.

The well established technique of reverberation mapping, for example, allows the mass to be estimated by tracking variations in the continuum flux from the accretion disc and comparing them with variations in emission lines from the broad line region (BLR) surrounding it.  The delay between variations allows the size of the BLR to be determined, which can be used to get a virial mass estimate. This technique forms the basis for the mass determinations for many of the AGN in the sample.  Other techniques derived from this exploit empirical correlations between, for example, optical luminosity at $\mathrm{5100\AA}$ and the BLR radius.  The study of \cite{2004ApJ...613..682P} suggests that there is a systematic uncertainty of a factor of $\sim 3$ in mass estimates derived using the reverberation mapping method.  There are also potential systematics in the estimation of NLS1 masses, possibly due to the flattening and inclination to the line of sight of the broad-line region; this could cause NLS1 BH masses to be underestimated under the reverberation mapping method \citep{2006A&A...456...75C}.

\begin{figure}
   \includegraphics[width=4.3cm]{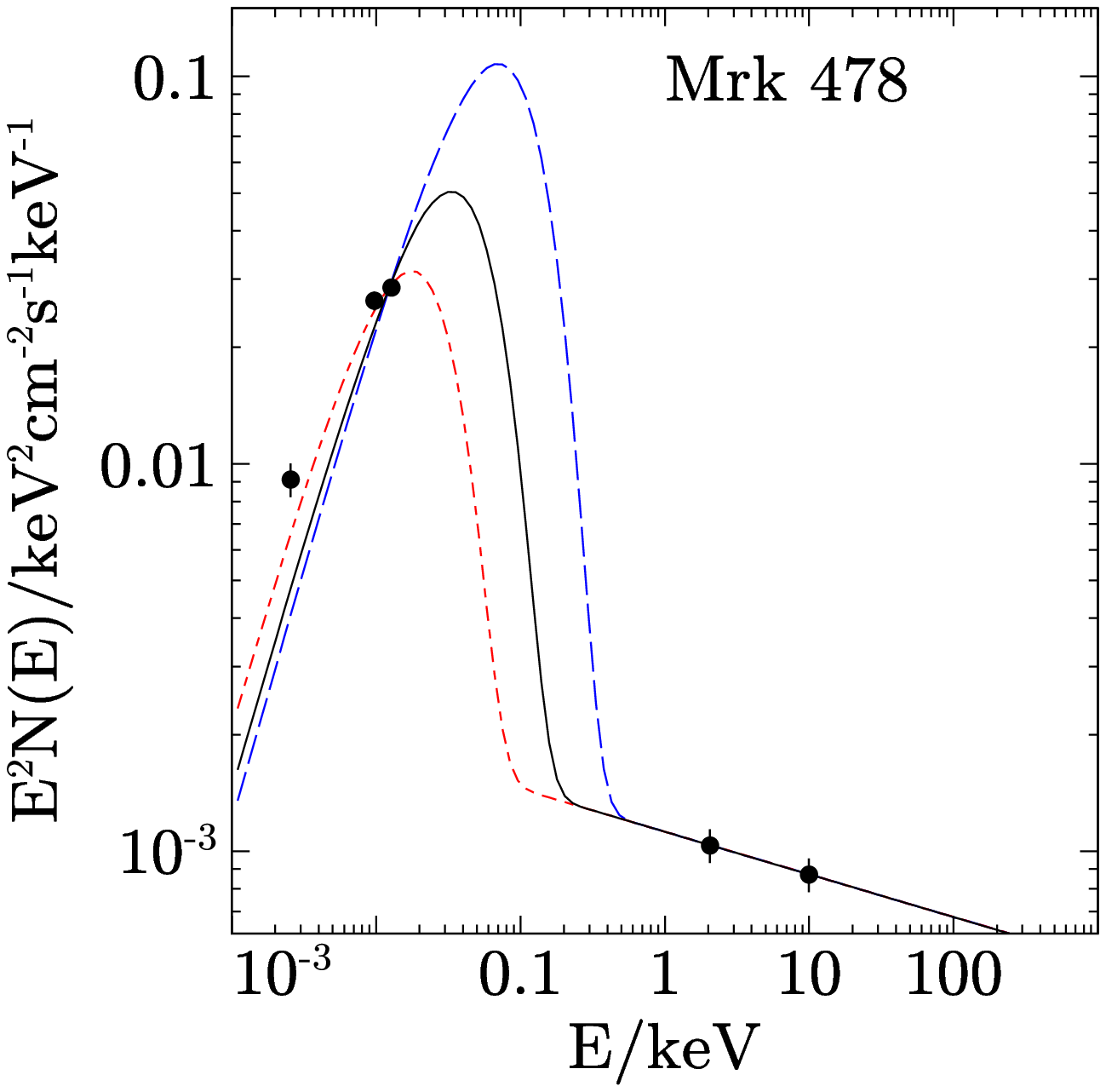}
   \includegraphics[width=4.3cm]{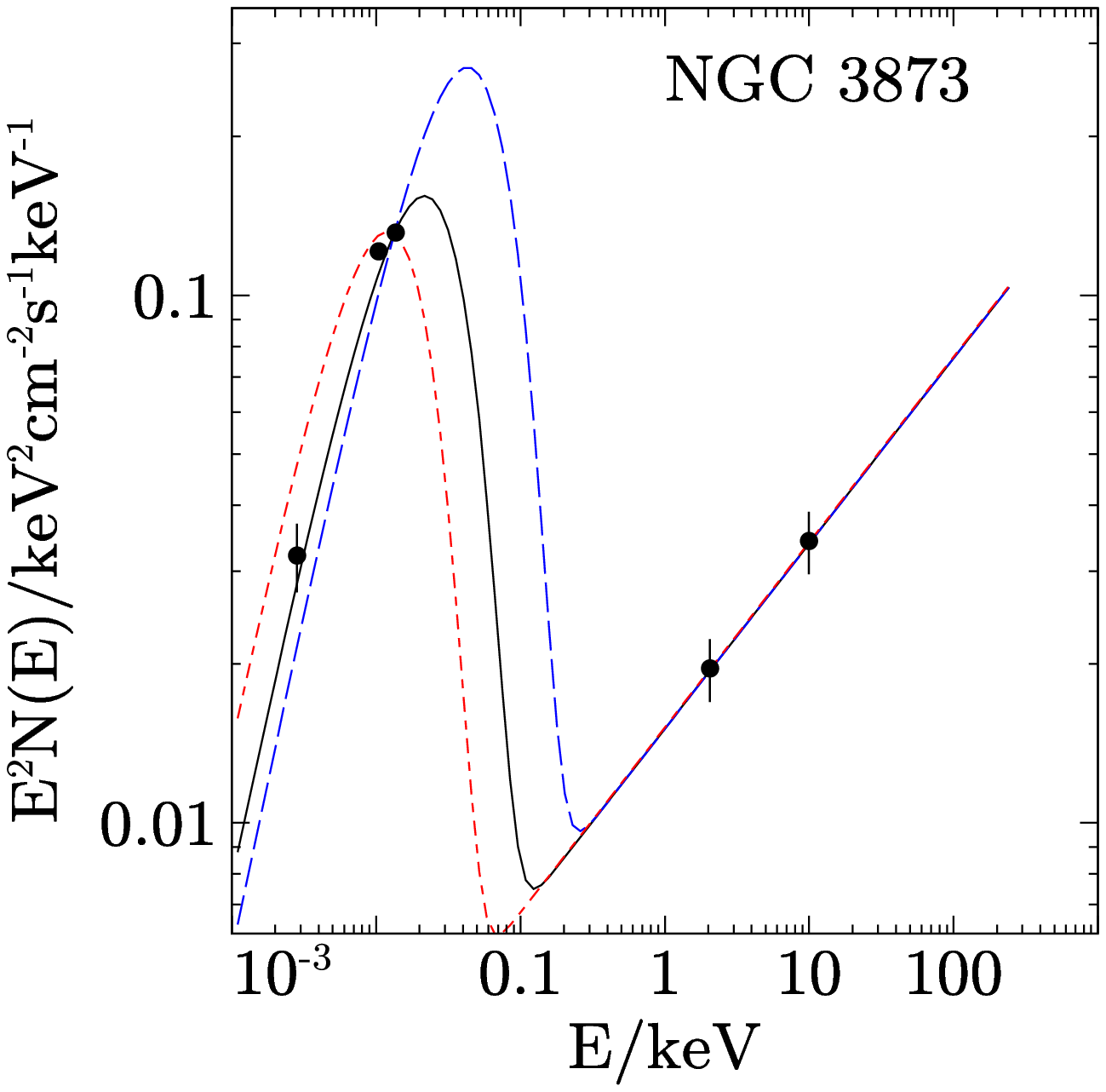}
   \caption{Effect of varying the SMBH mass on the SEDs generated.  Red dot-dashed lines show the model fit with mass increased by a factor of 3, blue dashed lines show the fit with the mass decreased by the same factor; other conventions as in Fig. \ref{sampleSEDs}.  For NGC 3873, increasing the mass decreases $\kappa_{\mathrm{2-10keV}}$ by 8 per cent and decreasing the mass increases it by 40 per cent.  For Mrk 478, increasing the mass reduces $\kappa_{\mathrm{2-10keV}}$ by 35 per cent and decreasing the mass increases it by 107 per cent.  These two objects show particularly large changes; the average variations were significantly lower.}
\label{massvar}
\end{figure}

We attempt to assess the importance of such variations by varying the SMBH masses up and down by a factor of 3, since this represents a typical systematic uncertainty in BH mass estimation methods.  We exclude MR 2251-178 since the mass estimate used is already an upper limit \citep{2002MNRAS.329..209M}.  A sample of SED fits generated with the varying BH masses are shown in Fig. \ref{massvar}.

We find that if the mass is increased by a factor of 3, the bolometric correction decreases by $14\pm13$ per cent on average, and decreasing the mass by the same factor increases the bolometric correction by $8\pm15$ per cent on average. If the fits produced from varying the mass produced reduced chi-squared values greater than 10 times higher than their previous values, we discarded these objects from the average since these are likely to represent particularly poor fits to the data giving unrealistic SED shapes.  The averages presented are of a similar order to the random error in bolometric corrections due to other sources, and so systematic uncertainties of this order (a factor $\sim 3$) in the mass are significant, but not more so than uncertainties from other sources.

\section{Parallels with Galactic Black Hole Accretion}

\label{GBHacc}

Our results may provide another perspective on the question of whether AGN accretion processes can be understood as a scaled-up version of Galactic Black Hole (GBH) accretion.  \cite{2006Natur.444..730M} discuss the importance of X-ray emission in this regard, since it emerges from very close to the black hole.  They specifically discuss the topic of a universal relation between black hole mass and `break timescale' (timescale at which power spectral density steepens) across BH masses ranging from Stellar to Supermassive. The review of \cite{2006ARA&A..44...49R} discusses the three state picture for GBH accretion, identifying the `hard' state (formerly known as the low/hard state), the thermal state (formerly the `high/soft' state) and the steep power-law (SPL) state (formerly the `very high' state); along with the preponderance for quasi-periodic oscillations in the thermal/SPL states.  The Eddington ratio $\mathrm{L_{bol}/L_{Edd}}$ is a useful separator of these states, and there is a progression from low to high Eddington ratios as these sources undergo transitions between the hard, thermal and SPL states.  The latter is often observed close to the Eddington limit.  In this section, we discuss our results in the context of the emergent unified pictures of accretion on all mass scales.

\subsection{Bolometric Correction as a function of Eddington Ratio}
\label{bcasafnofedd}

\begin{figure*}
   \includegraphics[width=10cm]{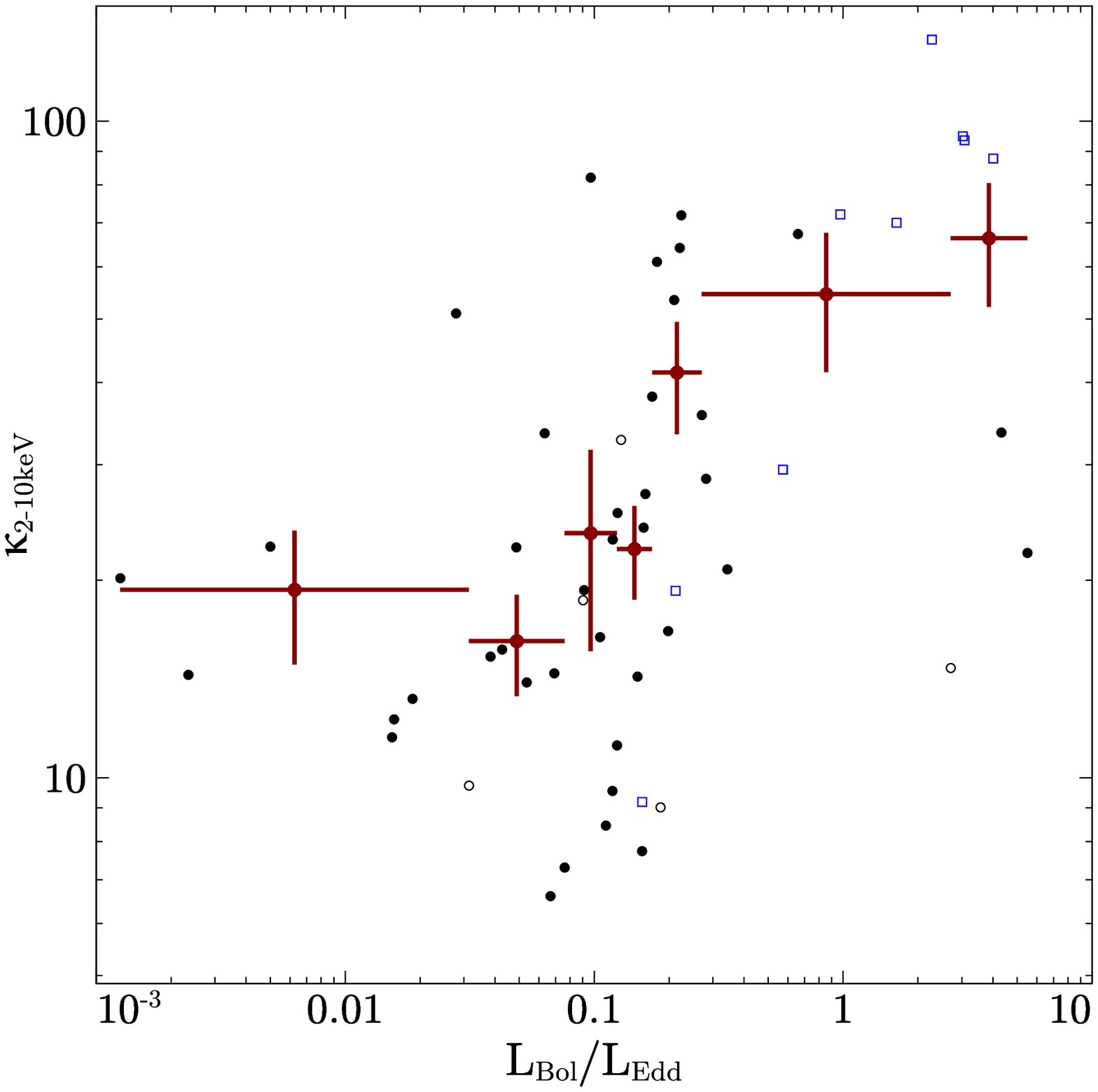}
   \includegraphics[width=10cm]{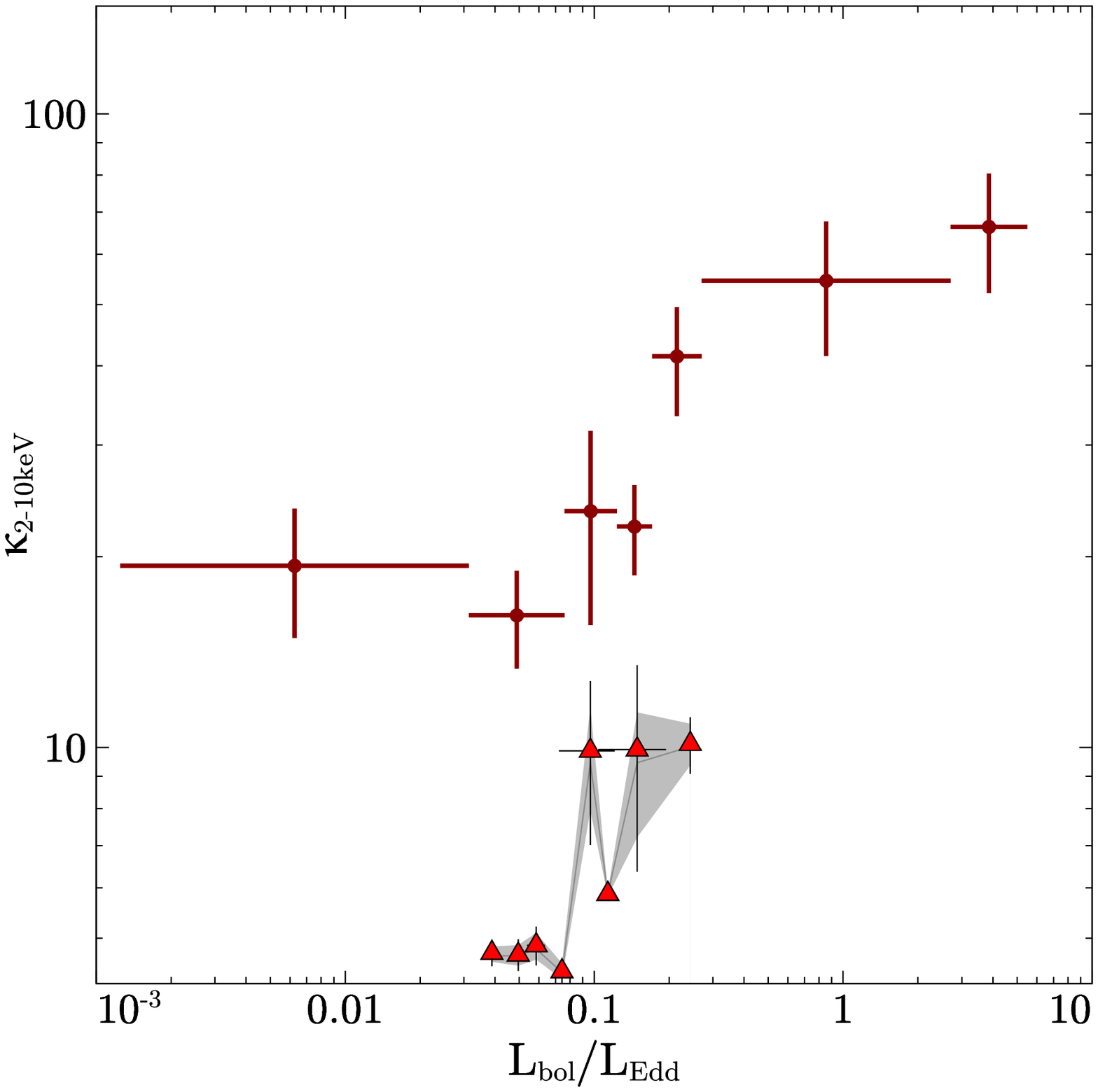}
   \caption{Upper panel: \210keV bolometric correction against Eddington ratio for the AGN in the sample.  Blue empty squares represent NLS1 nuclei, black empty circles represent radio loud AGN and dark red points with error bars are binned results.  Lower panel: bolometric corrections for the GBH GX 339-4 (red triangles) superimposed on binned AGN results. Individual AGN data points removed for clarity.  Grey shaded region shows the interquartile range calculated from Monte-Carlo realisations of the model fits, and the grey line shows the median. The possible problems due to the inclusion of $\mathrm{L_{bol}}$ on both axes are addressed in section \ref{syst_edd}.}
\label{bcvsledd}
\end{figure*}

We present the bolometric corrections for AGN plotted against Eddington ratio in Fig. \ref{bcvsledd}, top panel, with the outlier PG1011-040 (X-ray weak) removed.  The Eddington ratio can be understood as the luminosity scaled by the mass of the central black hole, and presented in this form there appears to be a distinct step change in bolometric correction at an Eddington ratio of $\sim$0.1. We confirm previous results which highlight the tendency for NLS1 nuclei to have high Eddington ratios  \citep{1995MNRAS.277L...5P}.   If accretion states in AGN are qualitatively similar to those in GBHs, the X-ray bolometric corrections could be a useful diagnostic of the accretion state in a black hole and the step-up in bolometric corrections above $\sim0.1$ Eddington could be indicative of AGN undergoing a transition between such accretion states.

\cite{2004ApJ...607L.107W} present the variation of $\mathrm{L_{2-10\mathrm{keV}}}/\mathrm{L_{bol}}$ (i.e. $1/\kappa_{2-10\mathrm{keV}}$) against Eddington ratio for a sample of AGN. Their their results agree qualitatively with ours; however, their bolometric luminosities are not always calculated from the SED continuum, as some of them are estimated using $\mathrm{L_{bol} \approx 9 \lambda L_{\lambda}(5100{\AA})}$ (\citealt{2000ApJ...533..631K}).  For those objects where bolometric luminosities are integrated from the multiwavelength continuum, they use values from \cite{2002ApJ...579..530W} and \cite{2004AJ....127..156G}, who do not consistently integrate over the same ranges (the former integrate from infrared to X-rays, whereas the latter integrate from optical to X-rays as we do here).  It is therefore possible that our approach of constructing SEDs afresh and integrating over the same energy range in each case may provide a window on some of the intricacies in the variation of the bolometric correction with Eddington ratio (such as the step change perceived here).

\begin{figure}
   \includegraphics[width=7.5cm]{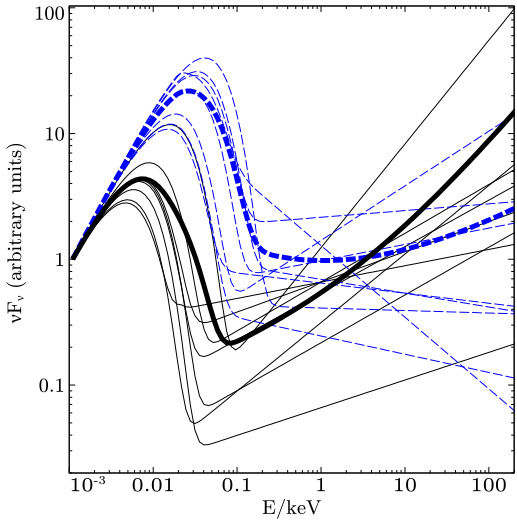}
   \caption{SEDs (normalised at 1 eV) for AGN with Eddington ratios in the range $0.0012 < \mathrm{L_{bol}/L_{Edd}} < 0.032$ (black solid curves) and $0.27 < \mathrm{L_{bol}/L_{Edd}} < 2.7$ (blue dashed curves).  The thicker curves show the mean SEDs calculated from the two sets of SEDs.}
\label{meanSEDhighlowEdd}
\end{figure}

\subsubsection{Accretion States}
Previous work has allowed the characterisation of distinct spectral features in the different accretion states of GBH binaries.  At low Eddington ratios, the low/hard state is characterised by a hard ($\mathrm{\Gamma < 2}$) power law, often dominating the black body disk spectrum.  In the high/soft state, the power law spectrum is much softer ($\mathrm{\Gamma > 2}$), but often the power law again dominates over the disk component.  In intermediate states such as the thermal state, the black body is comparatively more prominent.  There is a progression for increasing overall flux in the various components going from low to high states.  To contrast these characteristics with the AGN SED at high and low Eddington ratios, we reproduce the SEDs normalised at 0.1 eV, for eight AGN in each of the bins for Eddington ratios $0.0012 < \mathrm{L_{bol}/L_{Edd}} < 0.032$ (bin 1) and $0.27 < \mathrm{L_{bol}/L_{Edd}} < 2.7$ (bin 2) in Fig. \ref{meanSEDhighlowEdd} (we do not use the highest Eddington ratio bin since uncertainties surrounding mass determinations could be artificially producing extremely super-Eddington accretion rates).

The average SEDs for the high and low Eddington AGN allow us to draw some parallels with high and low accretion states in GBH binaries.  The lower Eddington AGN have a harder spectral shape in X-rays, as expected for the low/hard state, and the higher Eddington AGN have softer X-ray spectra as expected of the high state.  The dependence of the X-ray photon index on Eddington ratio in AGN has been commented on by other authors including \cite{2006ApJ...646L..29S}, who find a correlation between $\mathrm{\Gamma}$ and $\mathrm{L_{bol}/L_{Edd}}$.  Their findings are broadly echoed by our average SEDs.  In this study, we are also able to see how optical--UV emission varies with Eddington ratio. We find that the shift in the peak energy of the SED is roughly that expected from the dependence of Eddington ratio on temperature;  For a disk the black body luminosity scales as $L \propto T^{4}M^{2}$, and since $L_{Edd} \propto M$,  The ratio $L/L_{Edd}$ will scale as $T^{4}M$.  The two bins used span a factor of $\sim140$ in $\mathrm{L_{bol}/L_{Edd}}$; the expected range in temperatures would therefore be a factor of $\sim3.4$, assuming a similar distribution of masses for the AGN in the two bins.  The ratio of temperatures observed between the two bins is indeed $\sim3.4$.   The assumption of similar mass distributions does appear to be valid here: the average values of $\mathrm{log(M_{BH}/M_{\odot})}$ in the low and high Eddington ratio bins are $7.34\pm0.38$ and $7.76\pm0.19$ respectively.  The similar mass distributions seen in the two bins may imply that the disk structure is probably similar in the `hard' and `soft' states, and is broadly consistent with the hypothesis that the disk extends inwards close to the black hole without truncation.

\subsubsection{Coronal Fraction}
We can use the \210keV bolometric correction to calculate the total coronal fraction for our two fiducial high and low Eddington AGN SEDs, by extrapolating the coronal emission over the whole X-ray energy range.  We employ the mean photon indices for high and low Eddington ratios to calculate the factor difference between the 0.1--250 keV flux and the \210keV flux in the two states.  We can then estimate the power-law fraction, or `coronal fraction', using typical bolometric corrections of $\sim 20$ for low Eddington and $\sim 50$ for high Eddington.  The observed coronal fractions emerge as $\sim 35$ per cent for low Eddington ratio and $\sim 10$ per cent for high Eddington ratio, representing a fall of almost a factor of 4 between low and high Eddington ratio regimes.

However, the presence of X-ray reflection introduces some subtle modifications to this picture   We have so far assumed that the X-ray emission represents the total coronal emission $\mathrm{L_{C}}$ and the UV bump ($\mathrm{L_{D}}$)is solely from the accretion disk.  If reflection is taken into account, the X-rays we see are in fact only part of the total coronal emission (assumed to be 1/2 for simplicity here, corresponding to a reflection fraction of unity), and the rest is radiated onto the disk. The UV bump then contains both emission from the thermal accretion disk and some fraction \emph{f} of reprocessed X-rays from the corona ($\mathrm{L_{reproc}=0.5fL_{C}}$).  The \emph{true} coronal fraction $\mathrm{F_{true} = L_{C}/(L_{C}+L_{D})}$ will then differ from the observed fraction $\mathrm{F_{obs} = 0.5 L_{C} / (0.5 L_{C} + L_{D} + 0.5 f L_{C})}$.  Using a simple \textsc{pexrav} reflection model in \textsc{xspec} with reflection fration as above (canonical abundances assumed), we obtain absorption fractions $\mathrm{f\sim2/3}$ and $\sim5/6$ for low and high Eddington photon indices respectively.  This results in true coronal fractions of $45$ per cent for low Eddington and $11$ per cent for high Eddington.

\subsubsection{Implications for Surveys}
\begin{figure}
   \includegraphics[width=7.5cm]{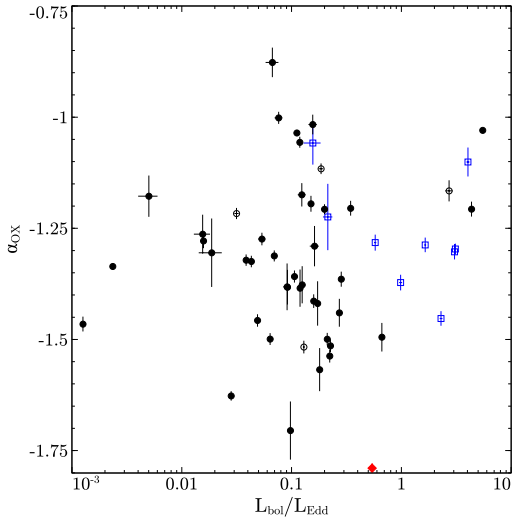}
   \caption{Plot of $\mathrm{\alpha_{OX}}$ against $\mathrm{L_{bol}/L_{Edd}}$ for the sample.  Key as in Fig. \ref{bcvslumallagn}.}
\label{alphaOXvsEddratio}
\end{figure}

\cite{2006ApJ...648..128K} find a dearth of AGN below $\mathrm{L_{bol}/L_{Edd}}\sim0.1$ in a joint X-ray -- IR -- optical study of the Bo\"otes field.  If objects are indeed missing then they would have masses of about $10^{9}M_{\odot}$, and so are not well represented by those in our study.  If there is an abrupt change in bolometric correction of these objects at an Eddington ratio of 0.1 such as found here, then this may influence their selection of objects.  Our results do not obviously account for their finding in a straightforward way, although we note that objects in the high and low Eddington ratio bins shown in Fig. \ref{meanSEDhighlowEdd} have very different ionizing photon fluxes (here taken to be the flux between 0.036 and 0.1 keV). Using the SEDs from Fig. \ref{meanSEDhighlowEdd}, we calculate that the higher $\mathrm{L_{bol}/L_{Edd}}$ AGN have $\sim40$ times the ionizing photon flux of the lower $\mathrm{L_{bol}/L_{Edd}}$ AGN, despite having similar optical to X-ray ratios ($\mathrm{\alpha_{OX}}$).  This could affect the observed line luminosities and hence also any selection criteria based on them.  We find further confirmation of this in Fig. \ref{alphaOXvsEddratio}, where we present $\mathrm{\alpha_{OX}}$ against $\mathrm{L_{bol}/L_{Edd}}$.  The relatively constant value of $\mathrm{\alpha_{OX}}$ seen across all Eddington ratios in the sample reiterates the point that the optical--X-ray ratios in AGN do not provide us with sufficient information on the important variation in the UV which is needed for calculating accurate bolometric corrections.  We see from Figs. \ref{meanSEDhighlowEdd} and \ref{bcvsledd} that this variation is significant.

\begin{figure}
   \includegraphics[width=7.5cm]{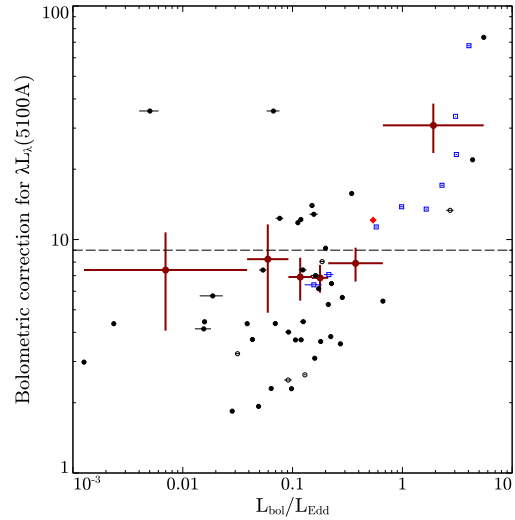}
   \caption{Optical bolometric correction $\mathrm{L_{bol}/{\lambda}L_{\lambda}(5100\AA)}$ against Eddington ratio $\mathrm{L_{bol}/L_{Edd}}$. Key as in Fig. \ref{bcvsledd} (the red diamonds represent X-ray weak objects as in previous figures). The dotted line shows the bolometric correction of $9$ \protect\citep{2000ApJ...533..631K} which is often used to estimate bolometric luminosities from monochromatic optical luminosities.}
\label{optbolcorvseddratio}
\end{figure}

We can briefly comment on the traditional estimation technique of using $\mathrm{9{\lambda}L_{\lambda}(5100\AA)}$ to estimate the bolometric luminosity, taken from \cite{2000ApJ...533..631K}.  The optical bolometric correction $\mathrm{L_{bol}/{\lambda}L_{\lambda}(5100\AA)}$ is plotted against $\mathrm{L_{bol}/L_{Edd}}$ in Fig. \ref{optbolcorvseddratio}.  We notice that for low Eddington AGN, the bolometric correction of $9$ (with a scatter comparable to the factor $\sim2 $ identified by \citealt{2006ApJS..166..470R}) is relatively accurate, but deviates from this at high Eddington ratios.  The optical bolometric correction shows a similar trend to the X-ray bolometric correction, increasing with Eddington ratio. Deviations from the traditional estimation technique may therefore be particularly important for NLS1s, if indeed their super-Eddington accretion rates are real.  However, since the optical points may be prone to some uncertainties such as contamination from host galaxy starlight etc., we caution against extrapolating too much from this plot, apart from the general trend for increasing bolometric correction with Eddington ratio.

\subsubsection{Consideration of Systematics}
\label{syst_edd}

It is important to note that where bolometric correction is plotted against Eddington ratio, there is the potential to introduce false correlations since $\mathrm{L_{bol}}$ is present on both axes.  However, if we generate a simple scatter plot using randomly generated input parameters taking values in the ranges available in our data, we find the slope of any implied or `false' trend is far steeper than the trend seen here, leading us to believe that the trends we see are real.  However the lack of objects in the lower-right regions of the bolometric correction against $\mathrm{L_{bol}/L_{Edd}}$ plots could be in part due to this.  There is also the issue of the preponderance for high mass objects to have high X-ray luminosities.  Neither of these systematics would reproduce the `step' change in bolometric correction seen here, and so we propose that these systematics are not immediately relevant.  

Finally, we note the possibility that the selection bias inherent in the \emph{FUSE} sample could be artificially producing the step-change seen here at $\sim0.1$ Eddington.  The AGN represented in the \emph{FUSE} sample are relatively UV bright and also low in redshift.  The large number of AGN clustering at Eddington ratios of $\sim0.1$ could, for example, reflect a tendency for local AGN to have this accretion rate, explaining in part the lack of significant numbers of low Eddington AGN.  The UV brightness of the sample would reinforce this, implying a bias away from `low state' AGN, which are expected to be UV-faint (Figure \ref{meanSEDhighlowEdd}).  The domination of NLS1 nuclei at high Eddington may also not be representative of the populations making up the XRB.  However confirmation or refutation of this will have to wait for a more extensive study including AGN at higher redshifts.

\subsection{Comparison with GX 339-4}

In order to compare the bolometric corrections between AGN and GBHs,
we need to bear in mind how different components of the emission scale
with mass; the scaling of the disk temperature with mass causes the
`BBB' in the UV arising from the disk emission in AGN to be shifted to
X-ray energies in GBHs.  However, the power-law component arising from
the coronal emission is scale-independent.  So to estimate the total
accretion flux in GBHs we evaluate the flux from the lowest energy of
the disk bump until $\sim\mathrm{100keV}$, and then to evaluate the
\210keV power-law flux we neglect the disk component to
obtain only the coronal flux in \210keV.  We use fits to
RXTE, Integral and BeppoSAX data on the GBH GX 339-4, provided by
\cite{2006ApJ...653..525M}, \cite{2003A&A...408..347C} (hard state),
\cite{2006MNRAS.367.1113B} (soft intermediate/thermal state) and
\cite{2004ApJ...606L.131M} (SPL state) to calculate bolometric
corrections.  The \210keV bolometric corrections against
Eddington ratio are presented in the lower panel of Fig.
\ref{bcvsledd}.

It is interesting to note that the bolometric corrections do increase as expected as the source moves between the hard, thermal and SPL states (increasing Eddington ratio).  The transitional region between the hard and SPL states is at around $\mathrm{L_{bol}/L_{Edd}}\sim0.1$, allowing us to tentatively identify the high and low bolometric correction regions seen in AGN as being from two relatively distinct accretion states.  It is known from the literature, however, that a number of GBH binaries undergo state transitions at significantly lower Eddington ratios, of the order $\sim2$ per cent Eddington \citep{2003A&A...409..697M}.  They discuss the variation in transition luminosity in going from hard--to--soft and soft--to--hard states, and the hysteresis implied by it.  It is possible that we may be witnessing similar hysteresis in our AGN sample by virtue of the lack of a very sharp, `thin' transition between low and high bolometric corrections around $\mathrm{L_{bol}/L_{Edd}\sim0.1}$.

\begin{figure}
  \includegraphics[width=8.8cm]{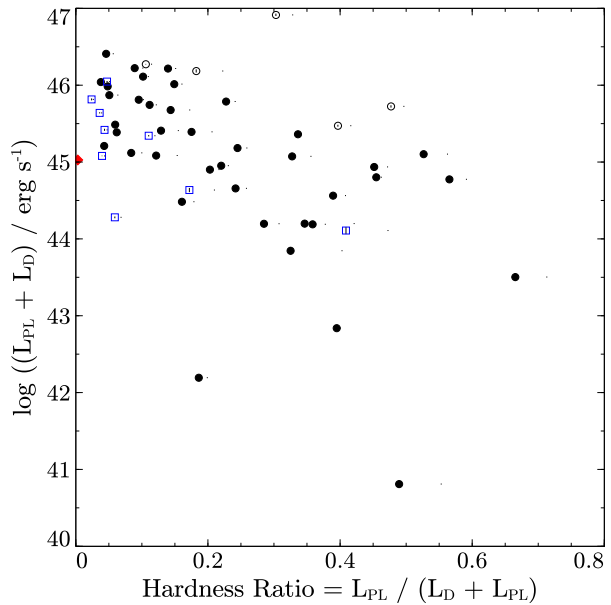}
   \caption{HID/DFLD for the AGN in our sample.  Key as in Fig. \ref{bcvsledd}.}
\label{HID}
\end{figure}

The work of \cite{2006MNRAS.372.1366K} discusses the use of `Hardness--Intensity' or `Disk Fraction Luminosity' diagrams (HIDs or DFLDs) for characterising the accretion states in both AGN and X-ray binaries.  To construct the AGN HID, They define the `power-law luminosity' $\mathrm{L_{PL}}$ as $3\times\mathrm{L_{2-10keV}}$ and calculated the disk luminosity $\mathrm{L_{D}}$ by applying a bolometric correction from the \emph{B}-band.  The Hardness Ratio (HR) is defined as $\mathrm{L_{PL} / (L_{D}+L_{PL})}$.  The HID consists of $\mathrm{L_{D}+L_{PL}}$ plotted against the HR.  We attempt to construct a HID for our sample, but here we make use of the UV SED fit to work out the disk luminosity directly, defining $\mathrm{L_{D} = L_{0.001-0.1 keV}}$.  The resulting HID is presented in Figure \ref{HID}.

We note firstly that there are no objects in our sample with $0.8 < HR < 1.0$.  The lack of a significant `hard' population means that we can only make a limited comparison with their work using our sample.  However, we note that the high Eddington NLS1 nuclei are clustered in the region expected for soft-state objects, and the radio loud objects also lie where expected as given by Figure 11 of \cite{2006MNRAS.372.1366K}.
 
GBH binaries show striking changes in radio flux between hard and soft states (e.g. \citealt{2004ARA&A..42..317F}).  The radio behaviour of AGN is less clear.  There have been a number of recent studies of the radio loudness of AGN versus Eddington ratio (e.g. \citealt{2006astro.ph..4095S}, \citealt{2007astro.ph..1546P}, \citealt{2007astro.ph..2292M}) which show that higher radio loudness correlates with with lower Eddington fractions, but no sharp switch is seen.  Using 5 GHz radio luminosities from NED and the study of \cite{2003MNRAS.345.1057M}, we find that in our small sub-sample of 16 AGN in bins 1 and 2, there is no identifiable tendency for these low Eddington AGN to be radio loud (using the ratio of radio to X-ray luminosity as a measure of radio loudness).  A recent study by  \cite{2007astro.ph..2684X} discusses the possibility that even the correlations seen in GBH binaries may be influenced by significant systematics.  The radio -- X-ray connection in BH accretion therefore still requires clarification.

\section{Summary and Conclusions}

We have demonstrated that the hard X-ray bolometric correction for AGN may vary significantly from object to object, especially when considering classes of AGN such as Narrow Line Seyfert 1 nuclei and Radio Loud AGN.  Removing these classes of objects yields only a weak trend of increasing bolometric correction with luminosity (Fig. \ref{bcvslumallagnblacklistremoved}).  The uncertainties introduced by the lack of simultaneous SEDs, X-ray and systematics in SMBH mass estimation methods are also significant.  We argue that the dependences of $\kappa_{\mathrm{2-10keV}}$ on $\mathrm{L_{2-10keV}}$ proposed by \cite{2004MNRAS.351..169M} and \cite{2006astro.ph..5678H} do not reflect the spread observed in the real AGN population because they primarily take into account the SED shape via the $\mathrm{\alpha_{OX}-L_{\nu}(2500\AA)}$ correlation, missing the intrinsic spectral differences attributable to mass and luminosity seen at the peak of the BBB at around $\sim1000\mathrm{\AA}$.  We show that differences are important and are the main cause of the spread in bolometric corrections. 

At first sight this could be accounted for by invoking intrinsic reddening in UV-optical part of the SED, but the available information on reddening seems to suggest that this will increase bolometric corrections by at factors of $\sim 2$ at most, without necessarily aligning the bolometric corrections with the suggested forms.  There may also be a luminosity dependence of intrinsic reddening which could go against the trend for increasing $\kappa_{\mathrm{2-10keV}}$ with luminosity.  An understanding of intrinsic extinction therefore remains of paramount importance and further study into its effects in all types of AGN would be extremely valuable.  However, we establish that the effect of any of reddening corrections will be to increase the X-ray bolometric corrections.  In order to satisfy the Soltan argument, this would require an increase in the mass--to--radiation conversion efficiency, $\eta$.  Higher efficiences could potentially imply more rapidly spinning black holes and de-emphasize the relative importance of radiatively inefficient modes of accretion in the X-ray Background.

There are some potential limitations to this study due to the SED construction method.  These findings rely on the fitting of multiband data to a simple black-body disk model and power law.  In order to get a more robust and physically realistic SED and to improve some of the poor fits, more advanced models such as those discussed by \cite{2001ApJ...563..560B} for 3C 273 could be employed.  The use of only four to eight data points may not sufficiently constrain the SED fit to give an accurate bolometric correction, and the current method may miss out potentially important contributions to the accretion luminosity, such as the soft X-ray excess.  However the good agreement between $\mathrm{\alpha_{OX}}$ values from our SEDs and the sample of \cite{2005AJ....130..387S} would imply a sensible reconstruction of the continuum SED shape for most of the AGN in this sample.  Additionally, there is still much uncertainty as to how well the current disk models fit to the UV disk emission.  With the absence of any clear picture as to which is the `correct' model to be used for the AGN disk emission, the \textsc{diskpn} model used here provides a sensible first order approach to estimating the UV emission.  It is also clear that the extreme bolometric corrections (both high and low) seen in our sample are independent of such effects and so the (hitherto unseen) degree of spread in $\kappa_{\mathrm{2-10keV}}$ seen in our results can be confirmed as real.

While the relation between bolometric corrections and luminosities in the \210keV band seems subject to significant uncertainties, it appears that Eddington ratio is a better discriminator of populations of AGN with high and low bolometric corrections.  We find a distinct tendency for bolometric corrections of $15-25$ below a threshold of around $\mathrm{L_{bol}/L_{Edd}}\sim0.1$, with values of $40-70$ at higher Eddington ratios.  Making allowances for reflection, we find that this corresponds to a fall in the coronal fraction of a factor of $\sim 4$ when transitioning from low to high Eddington regimes.  The transitional region at $\mathrm{L_{bol}/L_{Edd}}\sim0.1$ is of particular interest because of the potential parallels to be drawn with GBH accretion states.  Notably, bolometric corrections for the GBH candidate GX 339-4 calculated from observations in the literature also display a transition at around $\mathrm{L_{bol}/L_{Edd}}\sim0.1$, marking the transitional region between hard and soft states. However, effects of hysteresis in state transitions on both SMBH and GBH scales could blur this simple picture.  Additionally, the aforementioned uncertainties due to intrinsic reddening, UV/optical and X-ray variability and systematic errors in mass estimates may again introduce uncertainties, and selection biases inherent in our sample must also be considered.

We note in passing that nearly one half ($\sim 45$ per cent) of the power of low Eddington ratio AGN appears to emerge in the corona, indicating that much of the accretion flow is extracted by magnetic fields and dissipated outside the disk (taking X-ray reflection into account).  For high Eddington ratio AGN, this fraction reduces to approximately one tenth.

The significant differences between bolometric corrections from non-simultaneous and simultaneous SEDs implies that to gain an accurate understanding of the AGN SED, a much larger catalogue of simultaneous SEDs akin to the sample of \cite{2006MNRAS.366..953B} is a pressing need. 

We finally note that we are presented with an opportunity to refine X-ray background modelling techniques further than has currently been done.  The pioneering studies of previous authors started off with an assumption of constant $\kappa_{\mathrm{2-10keV}}$ (e.g. \citealt{2002ApJ...565L..75E}, \citealt{1999MNRAS.303L..34F}), followed by the studies of \cite{Fabian:2003cz} and \cite{2004MNRAS.351..169M} who pointed out the importance of taking into account a luminosity dependence.  In particular, the latter study illustrates how using a physically realistic dependence of $\kappa_{\mathrm{2-10keV}}$ on luminosity yields a SMBH mass density from the X-ray background which agrees with that determined by matching the local and relic SMBH mass functions.  This work suggests that the more fundamental dependence may be on Eddington ratio, and a re-formulating the modelling process to incorporate this dependence may be useful for future studies.

\section{Acknowledgements}

RV acknowledges support from the Science and Technology Facilities Council (STFC - formerly PPARC) and ACF thanks the Royal Society for Support. We thank Poshak Gandhi for useful comments and discussions on this work.  We thank Jon Miller for useful discussions on studies of X-ray binaries in the literature.  We also thank the anonymous referee for helpful suggestions.  This research has made use of the Tartarus (Version 3.2) database, created by Paul O'Neill and Kirpal Nandra at Imperial College London, and Jane Turner at NASA/GSFC. Tartarus is supported by funding from PPARC, and NASA grants NAG5-7385 and NAG5-7067.  This research has also made use of the NASA/IPAC Extragalactic Database (NED) which is operated by the Jet Propulsion Laboratory, California Institute of Technology, under contract with the National Aeronautics and Space Administration.

\bibliographystyle{mnras} 
\bibliography{bolometric}

\newpage
\newpage
\begin{appendix}
\appendix
\section{Table of Data Sources}
\label{appendix1}

\begin{table*}

\begin{tabular}{p{2.5cm}p{1.5cm}p{3.7cm}p{2.5cm}p{3cm}p{2.5cm}}
\hline
AGN name&Redshift&Optical data source&UV data source&X-ray data source&Mass data source\\\hline
\hline

NGC 4395&0.001064&(None found)&[1]&ASCA&[2]\\
NGC 3227&0.003859&HST [3]&HST--STIS [3]&ASCA&[4]\\
NGC 6814&0.006&Hale telescope ([5], [6])&IUE [5]&ASCA&[7]\\
NGC 3516&0.008836&HST [8]&HST--STIS [8]&ASCA&[4]\\
NGC 4593&0.009&(None found)&IUE [9]&ASCA&[4]\\
NGC 3873&0.01&[10]&FUSE&ASCA&[11]\\
NGC 7469&0.016&Mt. Wilson telescope [12]&FUSE&ASCA&[11]\\
Mrk 79&0.022&(None used -- [10] available)&FUSE&XMM [13]&[11]\\
Ark 564&0.025&(None used -- [10] available)&FUSE&ASCA&[14]\\
Mrk 335&0.026&[15]&FUSE&ASCA&[11]\\
Mrk 279&0.03&[10]&FUSE&ASCA&[16]\\
Mrk 290&0.03&HST and KPNO [17]&FUSE&ASCA&[15]\\
Ark 120&0.032&[10]&FUSE&ASCA&[11]\\
Mrk 509&0.034&HST and KPNO [17]&FUSE&ASCA&[11]\\
ESO141-G55&0.036&[10]&FUSE&ASCA&[11]\\
KUG 1031+398&0.042&(None found)&FUSE&ASCA&[18]\\
NGC 985&0.042&[10]&FUSE&ASCA&[4]\\
Mrk 506&0.043&HST and KPNO [17]&FUSE&ASCA&[19]\\
Fairall 9&0.047&(None used -- [10] available)&FUSE&ASCA&[11]\\
3C 382&0.058&(None used -- [20] available)&FUSE&ASCA&[21]\\
PG 1011-040&0.058&[15]&FUSE&ASCA&[15]\\
I Zw 1&0.061&[15]&FUSE&ASCA&[15]\\
TON S180&0.062&UK Schmidt telescope [22]&FUSE&ASCA&[4]\\
II Zw 136&0.063&[15]&FUSE&Ginga [23]&[15]\\
PG 1229+204&0.063&[15]&FUSE&ROSAT [24]&[15]\\
TON 951&0.064&HST and KPNO [17]&FUSE&XMM [25]&[15]\\
MR 2251-178&0.066& [26]&FUSE&Chandra [27]&[28]\\
Mrk 205&0.071&(None used -- [26] available)&FUSE&ASCA&[29]\\
Mrk 478&0.079&[15]&FUSE&ASCA&[15]\\
PG 1211+143&0.081&[15]&FUSE&ASCA&[11]\\
Mrk 1383&0.086&[15]&FUSE&ASCA&[15]\\
PG 0804+761&0.1&[15]&FUSE&ASCA&[11]\\
PG 1415+451&0.114&[15]&FUSE&XMM [30]&[15]\\
Mrk 876&0.129&[15]&FUSE&XMM [30]&[15]\\
PG 1626+554&0.133&[15]&FUSE&XMM [30]&[15]\\
PG 0026+129&0.142&[15]&FUSE&ASCA&[11]\\
PG 1114+445&0.144&[15]&FUSE&XMM [30]&[15]\\
PG 1352+183&0.152&[15]&FUSE&XMM [30]&[15]\\
PG 1115+407&0.154&[15]&FUSE&XMM [30]&[15]\\
PG 0052+251&0.155&HST and KPNO [17]&FUSE&ASCA [31]&[15]\\
PG 1307+085&0.155&[15]&FUSE&XMM [30]&[11]\\
3C 273&0.158&HST and KPNO [17]&FUSE&ASCA&[11]\\
PG 1402+261&0.164&Univ. of Hawaii 2.2m [32]&FUSE&ASCA&[33]\\
PG1048+342&0.167&[15]&FUSE&XMM [30]&[15]\\
PG 1322+659&0.168&HST and KPNO [17]&FUSE&XMM [34]&[15]\\
PG 2349-014&0.174&HST and KPNO [17]&FUSE&ASCA&[33]\\
PG 1116+215&0.176&[15]&FUSE&ASCA&[33]\\
4C +34.47&0.206&HST and KPNO [17]&FUSE&XMM [35]&[36]\\
PG 0947+396&0.206&HST and KPNO [17]&FUSE&XMM [30]&[15]\\
PG 0953+414&0.243&HST and KPNO [17]&FUSE&ASCA&[15]\\
PG 1444+407&0.267&[15]&FUSE&ASCA&[33]\\
PG 1100+772&0.311&HST and KPNO [17]&FUSE&ASCA&[15]\\
PG 1216+069&0.331&[15]&FUSE&XMM [30]&[15]\\
PG 1512+370&0.371&[15]&FUSE&XMM [30]&[15]\\
\hline

\end{tabular}
\label{table:datasources}
\caption{Data sources for Optical, UV, X-ray and SMBH mass estimates for AGN in the sample.  FUSE observations are all taken from \protect\cite{2004ApJ...615..135S} and ASCA observations are from the \textsc{Tartarus} database, unless otherwise stated. Other references: [1] - \protect\cite{2004ApJ...612..152C}, [2] - \protect\cite{2005ApJ...632..799P}, [3] - \protect\cite{2001ApJ...555..633C}, [4] - \protect\cite{2005MNRAS.358.1405O}, [5] - \protect\cite{1995AcA....45..623C}, [6] - \protect\cite{1978AJ.....83.1257D}, [7] - \protect\cite{2001ApJ...553..677L}, [8] - \protect\cite{2000ApJ...534..180E}, [9] - \protect\cite{1994MNRAS.270..580S}, [10] - \protect\cite{1991trcb.book.....D}, [11] - \protect\cite{2000ApJ...533..631K}, [12] - \protect\cite{1970ApJ...162..743A}, [13] - \protect\cite{2005MNRAS.363...64G}, [14] - \protect\cite{2004ApJ...602..635R}, [15] - \protect\cite{2005MNRAS.356.1029B}. [16] - \protect\cite{2004ApJ...613..682P}, [17] - \protect\cite{2005ApJ...619...41S}, [18] - \protect\cite{2004MNRAS.352..823B}, [19] - \protect\cite{1983AcApS...3..113L}, [20] - \protect\cite{1989ApJS...69..365S}, [21] - \protect\cite{2004ApJ...614...91W}, [22] - \protect\cite{1990MNRAS.243..692M}, [23] - \protect\cite{1997MNRAS.288..920L}, [24] - \protect\cite{1996MNRAS.282..493C}, [25] - \protect{\protect\cite{2003A&A...398...81B}}, [26] - \protect\cite{1983ApJS...52..341M}, [27] - \protect\cite{2005ApJ...627...83G}, [28] - \protect\cite{2002MNRAS.329..209M}, [29] - \protect\cite{2005ApJ...626...89B}, [30] - \protect{\protect\cite{2005A&A...432...15P}}, [31] - \protect\cite{2005ApJS..161..185U}, [32] - \protect\cite{2001AJ....122.2791S}, [33] - \protect\cite{2001MNRAS.327..199M}, [34] - \protect\cite{2006MNRAS.366..953B}, [35] - \protect\cite{2004MNRAS.352..523P}, [36] - \protect\cite{2004ApJ...615L...9W}}
\end{table*}

\end{appendix}

\clearpage
\end{document}